\documentclass[twocolumn,aps,pra,nofootinbib,superscriptaddress,10pt,floatfix]{revtex4-2}
\usepackage[english]{babel}
\usepackage[T1]{fontenc}
\usepackage[utf8]{inputenc}
\usepackage{amsmath}
\usepackage{amssymb}
\usepackage{graphicx}
\usepackage{subcaption,physics}
\usepackage{braket}
\usepackage[bottom]{footmisc}
\usepackage{xcolor}
\usepackage{xurl}
\usepackage{array}
\usepackage[linktocpage]{hyperref}
\hypersetup{colorlinks=true,urlcolor=teal,citecolor=teal,linkcolor=teal}
\usepackage{multirow}
\usepackage{lipsum}
\usepackage{tikz}
\usepackage{dsfont}
\usepackage{csquotes}
\usepackage{nicefrac}
\usepackage{float}
\usepackage{siunitx}
\usepackage{bbold}
\usepackage{fixltx2e}

\usepackage[newcommands]{ragged2e}
\usepackage{graphicx,caption}
\usepackage[font=small,labelfont=bf,justification=justified]{caption}
\usepackage{booktabs}
\usepackage{array} 
\usepackage{multirow} 
\usepackage{float}
\usepackage{comment}
\usepackage{mdframed}
\usepackage{enumitem}
\usepackage{soul}
\allowdisplaybreaks
\usepackage[most]{tcolorbox}
\newtcolorbox[auto counter,number format=\Roman]{myalgobox}[2]{
  colback=white,
  colframe=gray!60!black,
  colbacktitle=gray!25,
  coltitle=black,
  title={Protocol~\thetcbcounter: #1},
  label={#2},
  boxrule=0.8pt,
  sharp corners,
  enhanced,
  attach boxed title to top center={yshift=-2mm},
  boxed title style={boxrule=0.5pt, colframe=gray!60!black},
  before skip=0pt,
  after skip=0pt,
}
\newcolumntype{M}[1]{>{\centering\arraybackslash}m{#1}} % centered column
\newcolumntype{L}{>{\raggedright\arraybackslash}p{4cm}} % left-aligned for quantity names
\usepackage[font=small]{caption}
\usepackage{ulem}

\definecolor{black}{rgb}{0.00,0.00,0.00}
\definecolor{cyan}{rgb}{0.00,0.60,0.70}
\definecolor{yellow}{rgb}{0.93,0.69,0.13}
\definecolor{magenta}{rgb}{0.80,0.20,0.50}
\definecolor{mintgreen}{HTML}{C5E7CD}
\definecolor{lightblue}{HTML}{9FD0E7}
\definecolor{brown}{rgb}{0.35, 0.20, 0.10}
\definecolor{skyblue}{rgb}{0.35, 0.20, 0.10}

\begin{document}

\title{Robust One-Sided Device-Independent Quantum Key Distribution\\ via High-Dimensional Steering}
\author{Monika Mothsara}
\affiliation{TU Wien, Atominstitut, Stadionallee 2, 1020 Vienna, Austria}
\author{Suraj Goel}
\affiliation{Institute of Photonics and Quantum Sciences, Heriot-Watt University, Edinburgh, UK}
\author{Bohnishikha Ghosh}
\affiliation{Institute of Photonics and Quantum Sciences, Heriot-Watt University, Edinburgh, UK}
\author{Vatshal Srivastav}
\affiliation{Institute of Photonics and Quantum Sciences, Heriot-Watt University, Edinburgh, UK}
\author{Will McCutcheon}
\affiliation{Institute of Photonics and Quantum Sciences, Heriot-Watt University, Edinburgh, UK}
\author{Mehul Malik}
\affiliation{Institute of Photonics and Quantum Sciences, Heriot-Watt University, Edinburgh, UK}
\author{Gl\'aucia Murta}
\affiliation{TU Wien, Atominstitut, Stadionallee 2, 1020 Vienna, Austria}

\begin{abstract}
Quantum key distribution (QKD) brings the promise of communication with information-theoretic security but is limited in practice due to its susceptibility to noise, losses, and device imperfections. To address these challenges, we propose a robust high-dimensional (HD) \textit{one-sided} device-independent QKD (1sDI-QKD) protocol and present a proof-of-principle experimental implementation using photons entangled in the transverse-spatial degree-of-freedom. 
We develop a systematic security analysis of HD 1sDI-QKD protocols, leveraging quantum steering to certify security, and evaluate achievable secret key rates for different measurement configurations and system dimensions using reverse reconciliation. Our analysis shows that increasing the dimension enhances robustness against both noise and loss.
We then demonstrate the key experimental building blocks required for implementing the protocol: (a) a high-quality source of high-dimensional photonic entanglement, and (b) a fully programmable, high-dimensional multi-outcome measurement device operating in up to dimension 11. Using these components, we obtain positive key rates for all investigated dimensions under the fair-sampling assumption, with the highest key rates achieved for dimension $d=7$. Finally, we discuss the steps required for a practical, loophole-free implementation of 1sDI-QKD in realistic regimes of loss and noise.
\end{abstract}

\maketitle

% SECTION I %%%%%%%%%%%%%%%%%%%%%%%%%%%%%%%%%%%%%%%%%%%%%%%%%%%%%%%%%%%%%%%%%%%%%%%%%%%%%%%%%%
\section{Introduction}

Quantum key distribution (QKD)~\cite{bennett2014quantum} is a cryptographic task that enables the establishment of a shared secret key by leveraging principles of quantum mechanics. Combined with a classical one-time pad scheme, QKD can lead to information-theoretically secure communication. As one of the most promising and successful applications of quantum information science, it is now commercially deployed in many countries worldwide \cite{Zhang2025-bo,Stanley_2022}. However, real-world implementations of QKD often face two major limitations: (i) noise and losses, which degrade protocol performance, and (ii) device imperfections, which are tied to the assumptions and rigor of the security proofs.

High-dimensional (HD) quantum systems offer a clear advantage in terms of information capacity and have been identified as a promising route to overcome the effects of noise and loss in entanglement distribution \cite{malik2026high,Erhard2020_highdimadvances,Cozzolino_highdimcomm}. In particular, qudit entanglement has been shown to be more resilient to losses compared to qubit entanglement, ranging from device-dependent to fully device-independent settings \cite{quditent_PhysRevX.9.041042,Zhu2021_pm,Designolle_highdimsteering, Vertesi2010bq,PhysRevX.12.041023,dekkers2025observinghighdimensionalbellinequality}. These benefits have also been predicted for QKD~\cite{sheridan_valerio_highdim,Gisin_dlevelsystems,doda2021quantum, Flo_highdim_prl,flo_highdim_finitesize}, with state-of-the-art experiments demonstrating that HD entanglement can enable higher secure key rates than those achieved with qubit entanglement~\cite{Lib:25, yu2025quantum, bulla2023nonlocal,doi:10.1126/sciadv.aee1333}. 

While HD variants of device-dependent QKD protocols can provide higher resilience to noise and loss, one has to work with the assumption that Alice and Bob have fully characterized measurement devices. On the other hand, device-independent (DI) QKD protocols can address device imperfections by ensuring security in the most adversarial setting, for example, where the devices are completely untrusted. However, existing security proofs have not yet shown any significant advantages associated with higher dimensions \cite{Rivera-Dean_2025}. Moreover, DI-QKD protocols are severely limited by losses~\cite{pironio2009device, Woodhead2021deviceindependent, Sekatski2021deviceindependent, Masini2022simplepractical}, restricting achievable distances \cite{Li_di2021exp, Liu_di2021exp, Lyndenshalm_2021exp}. 

\begin{figure}
         \centering
        \includegraphics[width=0.99\linewidth]{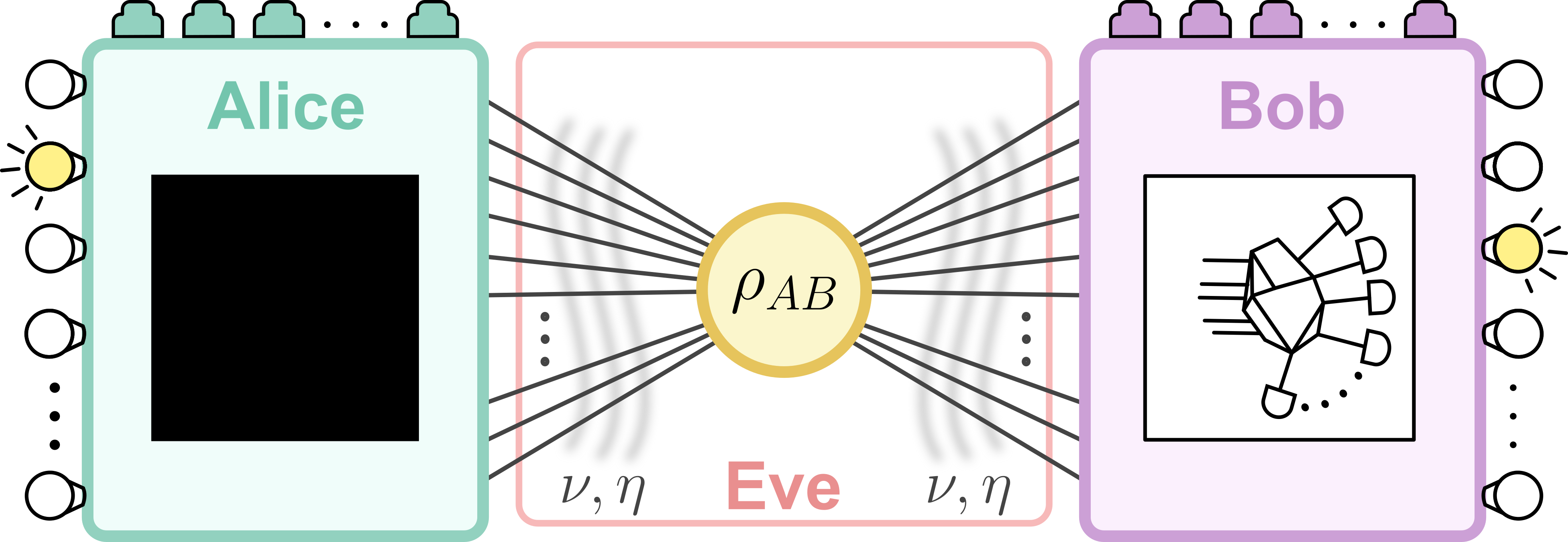}
        \caption{A schematic representation of a high-dimensional \textit{one-sided} device-independent QKD (1sDI-QKD) scenario. An untrusted source distributes bipartite entangled states to Alice, who holds an untrusted device, and Bob. Both parties perform high-dimensional multi-outcome measurements, and the resulting correlations enable the demonstration of quantum steering. Higher dimensions are expected to provide improved robustness against noise ($\nu$) and loss ($\eta$).}
        \label{HD1sdi_fig}
\end{figure}

An intermediate scenario is that of \textit{one-sided} device-independent QKD (1sDI-QKD) (see Fig.~\ref{HD1sdi_fig}) \cite{branciard2012one,masini2024onesideddiqkdsecurecoherent,roy2025secureonesideddeviceindependentquantum}, which provides a more feasible alternative compared to a fully DI-QKD framework. This approach is based on the notion of quantum steering, originally proposed by Schrodinger \cite{Schrodinger1935-hk} and later formalized by Wiseman \textit{et al.} \cite{wiseman2007steering}, and relies on the assumption that only one party's devices are trusted (say, Bob).  

1sDI-QKD is motivated by practical communication scenarios in which only one party has access to a trustworthy device. For example, a bank or data center may be able to deploy expensive trusted hardware while its customers rely on low-cost devices that may be insecure. A similar scenario may exist for secure communication with embassies or other facilities in conflict zones, where one may not have access to up-to-date certified hardware. Other examples include satellite-based communication, where maintaining onboard devices can be challenging. 

Despite recent experimental advances, HD-QKD protocols remain largely unexplored, particularly in the 1sDI setting. Existing security proofs for 1sDI-QKD protocols are predominantly limited to qubit systems or restricted measurement configurations \cite{branciard2012one,masini2024onesideddiqkdsecurecoherent,roy2025secureonesideddeviceindependentquantum}. At the same time, recent demonstrations of HD quantum steering in the spatial degree-of-freedom of light have shown increased robustness against noise and loss \cite{PhysRevX.12.041023}, highlighting the potential of this platform for implementing HD 1sDI-QKD.

A key challenge in extending QKD to higher dimensions lies in the implementation of genuine multi-outcome measurements.  Unlike qubit-based QKD demonstrations, where outcomes can be efficiently discriminated using off-the-shelf optical devices such as polarizing beam splitters, HD-QKD requires access to multi-outcome measurement devices capable of performing measurements in multiple HD bases. However, previous demonstrations of enhanced robustness for HD systems have largely relied on simulated multi-outcome measurements implemented via single-outcome detection schemes~\cite{quditent_PhysRevX.9.041042,Zhu2021_pm, designolle2021genuine} or binary-outcome configurations~\cite{PhysRevX.12.041023}, therefore limiting the applicability of promising entanglement distribution platforms for HD QKD.

Recent advances in wavefront shaping have led to the development of techniques for the control and manipulation of HD quantum states of light encoded in the transverse-spatial degree-of-freedom~\cite{goel2026quantum,malik2026high}, enabling HD quantum gates~\cite{brandt2020high,goel2024inverse,makowski2024large} as well as generalized multi-outcome measurements~\cite{goel2024inverse} 
These advances are fueled in part by devices such as multi-plane light converters~(MPLCs), which have recently enabled proof-of-concept device-dependent HD-QKD demonstrations~\cite{Lib:25}. 

In this work, we present a systematic security analysis of high-dimensional 1sDI-QKD protocols, where security is certified by quantum steering (see Fig.~\ref{HD1sdi_fig}). By relaxing assumptions on one of the parties while still leveraging the robustness of HD entanglement, this approach improves the practicality of fully-DI protocols while still allowing for partial device untrustworthiness. We show that reverse reconciliation exploits the inherent asymmetry of the steering scenario, leading to secret key rates and noise robustness that improve with increasing dimension.
We evaluate achievable secret key rates for protocols with different measurement configurations and system dimensions, deriving lower bounds on the visibility and detection efficiency required to generate secure keys in the asymptotic regime against collective attacks.

We then implement a proof-of-principle setup for HD 1sDI-QKD using pairs of photons entangled in the high-dimensional transverse spatial degree-of-freedom, distributed between two parties. In our setup, each party can employ an MPLC to realize multi-outcome measurements in all possible mutually unbiased bases~(MUBs) in dimensions up to $d=11$.
From the experimentally obtained data, we observe post-selected steering violations across all investigated dimensions, and we find that the corresponding secret key rate increases with dimension up to $d=7$. We further discuss routes towards detection-loophole-free implementations, highlighting the potential of our approach for real-world quantum communication scenarios.

% SECTION II %%%%%%%%%%%%%%%%%%%%%%%%%%%%%%%%%%%%%%%%%%%%%%%%%%%%%%%%%%%%%%%%%%%%%%%%%%%%%%%%%%
\section{1{s}DI-QKD scenario} \label{II}
Consider two spatially separated parties: Alice, whose measurement device is untrusted/uncharacterized, and Bob, whose device is fully characterized. An adversary, Eve, is assumed to have full control over the source and the channel, and to have manufactured Alice’s device. Eve's aim is to gain information about the secret key established between Alice and Bob. The two parties share a bipartite quantum state $\rho_{\mathrm{AB}}$, distributed by an untrusted source. Alice performs one of $m$ possible measurements, labeled by $x$ $\in$ $\{0, \ldots, \mbox{$m-1$}\}$, described by a set of (unknown) measurement operators $\{M_{a|x}\}_a$, each yielding one of the $d$ possible outcomes, $a$ $\in$ $\{0, \ldots, \mbox{$d-1$}\}$. Following Alice's measurement, Bob's system is steered into a conditional state $\rho_{a|x}$ with probability $p(a|x)$. Bob's information can be described by a set of unnormalized conditional states $\{\sigma_{a|x} = p(a|x)\rho_{a|x}\}_{a,x}$, denoted \textit{assemblage}. The elements of the assemblage are given by
\begin{equation}
    \sigma_{a|x} = \mathrm{tr}_A[(M_{a|x}\otimes\mathds{I}_{B})\rho_{\mathrm{AB}}]\,.
    \label{assemblage_B}
\end{equation} 
Note that $\mathrm{tr}[\sigma_{a|x}]=p(a|x)$, which corresponds to the probability that Alice obtains an outcome $a$ when performing measurement $x$. 

The EPR steering formalism asserts that Alice would not be able to convince Bob that she can steer his state if the observed \textit{assemblage} admits a local hidden state (LHS) model \cite{wiseman2007steering}. An LHS model is a classical model that corresponds to the situation where a source sends a classical hidden variable $\lambda$ to Alice with probability density $\xi(\lambda)$ and a fixed quantum state $\rho_{\lambda}$ to Bob. The variable $\lambda$ instructs Alice to output $a$ upon measuring $x$ with probability $p(a|x,\lambda)$. Consequently, the assemblage Bob observes has the form 
\begin{equation}
    \sigma_{a|x} = \int \mathrm{d} \lambda \xi(\lambda) p\left(a|x,\lambda\right)\rho_\lambda\,.
    \label{LHS_model}
\end{equation} 
An assemblage of this form is called unsteerable, and we denote it by $\{\sigma_{a|x}^{\mathrm{LHS}}\}_{a,x}$. A quantum state $\rho_{AB}$ is said to be steerable if there exist measurements that generate an assemblage that does not admit such a decomposition. 

To certify steering, Bob can test a steering inequality (SI) constructed from a set of Hermitian operators $\{N_{b|y}\}$ acting on a $d$-dimensional Hilbert space. These operators define a steering functional $\beta$, which can be written as
\begin{equation}
    \beta = \mathrm{tr} \sum_{a,x} N_{b=a|y=x} \sigma_{a|x} \leq \beta^{\mathrm{LHS}}\,,
    \label{S.I.}
\end{equation} 
where $\beta^{\mathrm{LHS}}$ is the maximum value attainable by all unsteerable assemblages, i.e., assemblages that admit an LHS model, Eq.~\eqref{LHS_model}. A violation of this bound, i.e., when $\beta > \beta^{\mathrm{LHS}}$, certifies the presence of steering and thereby rules out any possible local hidden state decomposition of the observed data~\cite{cavalcanti2016quantum,steering_review}.

In the presence of losses, if the data is post-selected, Alice and Bob may appear to demonstrate steering even when their shared correlations admit an LHS model. This gives rise to the so-called detection loophole, which occurs when the detected events do not constitute a fair sample of the entire ensemble.

Note that since Bob's measurement device is fully trusted and well-characterized, loss events on Bob’s side can be discarded without compromising security, as the fair-sampling assumption is justified~\cite{RevModPhys.81.1301}. In contrast, losses on Alice’s side cannot be treated in the same way, since an adversarial device could exploit biased detection to artificially produce steering violations. To avoid this loophole, Alice’s no-click events have to be explicitly included in the analysis, for example, as an additional measurement outcome. Within this framework, a general construction of loss-tolerant steering inequalities was introduced in~\cite{loss-tolerant_SI_Paul}, where the steering functional defined in Eq.~(\ref{S.I.}) is evaluated using the following operators
\begin{equation}
    N_{b|y}= 
    \begin{cases}
        N_{b|y} & \text { for } b=0, \dots, d-1 \\ 
        \alpha \mathds{I} & \text { for } b=\varnothing\,.
    \end{cases} \label{SIloss}
\end{equation}
where $N_{b|y}$ for $b\in\{0,1,\ldots, d-1\}$ denote the measurement operators corresponding to Bob’s trusted measurements, while the additional outcome, $b=\varnothing$,  accounts for Alice’s no-click events. The parameter $\alpha$ corresponds to the maximal overlap between any two of Bob's measurements, which in the case of MUBs equals $1/\sqrt{d}$.

In this work, we explore the asymmetry of the 1sDI scenario to propose protocols that employ a reverse reconciliation scheme and incorporate Alice’s extra-outcome strategy directly into the security analysis. We investigate HD 1sDI-QKD protocols using two complementary approaches for the security proof: the entropic uncertainty relation (EUR) framework~\cite{Berta2010-ls} and the estimation of conditional min-entropy certified by the violation of a steering inequality, in which case we use the inequality defined by the operators in Eq.~\eqref{SIloss}.

We restrict our analysis to asymptotic key rates under collective attacks, assuming independent and identically distributed (i.i.d.) rounds, with the aim of characterizing the advantages of high-dimensional systems. Recent techniques, such as the entropy accumulation theorem \cite{EAT_Dupuis2020-yq,rotem_DI_EAT}, provide a route to extending our analysis to finite-size key rates and security against coherent attacks.

% SECTION III %%%%%%%%%%%%%%%%%%%%%%%%%%%%%%%%%%%%%%%%%%%%%%%%%%%%%%%%%%%%%%%%%%%%%%%%%%%%%%%%%%
\section{Security proof and key rate} \label{III}
\begin{figure}[t!]
\begin{myalgobox}{\small 1sDI-QKD}{box:protocol1}
    \small 
    \begin{enumerate}[leftmargin=2pt]
        \item \textbf{For $i$ to $n$:}
        \begin{enumerate}[leftmargin=10pt,label=(\roman*)]
            \item A source (untrusted) distributes a bipartite entangled state $\rho_{\mathrm{AB}}$ to Alice (untrusted devices) and Bob (trusted devices). 
            \item Alice randomly chooses a measurement setting $x_i$ $\in$ $\{0,\ldots,\mbox{$m-1$}\}$ and records an outcome $a_i$ $\in$ $\{0,\ldots,\mbox{$d-1$},\varnothing \}$, where $\varnothing$ denotes a no-click event.
            \item Bob chooses his measurement setting $y_i$ $\in$ $\{0,\ldots,\mbox{$m-1$}\}$, measures the corresponding POVM $\{N_{b|y_i}\}_b$ and records his outcome $b_i$ $\in$ $\{0,\ldots,\mbox{$d-1$}\}$.
        \end{enumerate}
        \item \textbf{Sifting:} Alice and Bob reveal their choice of measurement bases publicly and keep the rounds where they match.
        \item \textbf{Parameter estimation \& Key generation:} Alice and Bob compare a small fraction of rounds to estimate their correlations across different measurement bases or, when applicable, to evaluate the violation of a steering inequality. The remaining rounds are used for key generation. Based on the observed correlations or steering violation, they decide whether to proceed or abort the protocol. We consider the following three protocol variants:
        \begin{enumerate}[leftmargin=10pt,label=\textbf{(\alph*)},ref=I\alph*]
            \item \label{protIa}
            \textbf{\textit{Two-basis protocol}:} One basis is used for key generation and correlation estimation, while the other basis is used solely for parameter estimation.
            \item \label{protIb}
            \textbf{{\itshape{\mbox{$d+1$}-basis} spot-checking protocol}:} One basis is designated for key generation, and all \mbox{$d+1$} bases are used to estimate the violation of a steering inequality.
            \item \label{protIc}
            \textbf{{\itshape{\mbox{$d+1$}-basis} with multiple key-generation basis protocol}:} The \mbox{$d+1$} bases are used for both key generation and to estimate the violation of a steering inequality.
        \end{enumerate}
        \noindent \item \textbf{Information reconciliation:} We employ reverse reconciliation to exploit the asymmetry advantage of the 1sDI setting, and therefore Alice (untrusted) corrects her key string according to Bob's (trusted).
        \item \textbf{Privacy amplification:} Alice and Bob perform privacy amplification to extract the final key.
    \end{enumerate}
\end{myalgobox}
\label{protocolbox}
\end{figure}
We consider the 1sDI-QKD protocol variants described in Protocol~\ref{box:protocol1}. To quantify their performance, we use the asymptotic secret key rate $r_{\infty}$, which measures the number of secret bits generated per round in the asymptotic limit of infinitely many rounds. For QKD protocols that use a one-way information reconciliation\footnote{In a one-way information reconciliation scheme, one of the parties, say Alice, keeps her raw string fixed, and information is exchanged for Bob to correct his raw string according to Alice's string.} scheme in Step 4 (see Protocol \ref{box:protocol1}), a lower bound is given by~\cite{devetak2005distillation}
\begin{equation}
    r_\infty^{\rightarrow} = H(A|E,X,Y) - H(A|B,X,Y),
    \label{keyrateAE}
\end{equation}
if Bob corrects his raw key according to Alice's string (direct reconciliation), or
\begin{equation}
    r_\infty^{\leftarrow} = H(B|E,X,Y) - H(B|A,X,Y)\,,
    \label{keyrateBE}
\end{equation}
if Alice corrects her raw key according to Bob's string (reverse reconciliation). Here $H(A|E, X, Y)$ ($H(B|E, X, Y)$) is the conditional entropy of Alice's (Bob's) measurement outcome conditional on the information available to Eve, quantifying Eve's uncertainty about Alice's (Bob's) outcome, and $H(A|B, X,Y)$ ($H(B|A, X,Y)$) measures the uncertainty Bob (Alice) has about Alice's (Bob's) outcome. $X$ and $Y$ denote the random variables associated with Alice's and Bob's measurement choices in a given round, respectively. We make them explicit in the key rate formula not only because they are publicly communicated in Step 2 of Protocol \ref{box:protocol1}, and therefore available to Alice, Bob, and Eve, but also because doing so allows us to clearly distinguish the key rate expressions corresponding to different protocol variants. The variables $A$ and $B$ denote the outcomes associated with measurement settings $X$ and $Y$, while $E$ represents the quantum system held by the eavesdropper. In QKD, Alice and Bob generate raw keys from their measurement outcomes, which may differ due to noise, loss, or other imperfections, and therefore require error correction, performed in Step 4 of Protocol~\ref{box:protocol1}. In this work, we focus on the reverse reconciliation strategy (see Section \ref{Reverse vs direct reconciliation} for more details). Therefore, unless explicitly stated otherwise, the key rates presented in this manuscript refer to eq.~\eqref{keyrateBE}, and the superscript $^{\leftarrow}$ will be omitted hereafter.

The term $H(B|A, X,Y)$ quantifies the amount of public information Bob must reveal during error correction for Alice to correct her key. This quantity can be estimated straightforwardly from the observed conditional probability distributions. In contrast, evaluating $H(B|E, X,Y)$ is more involved and generally requires numerical computation.

In the following, we analyze key rates for three high-dimensional 1sDI-QKD protocol variants. We first consider a \textit{two-basis} protocol, denoted Protocol~\ref{protIa}, in which Alice and Bob choose between two measurement bases in each round. We then study two protocols in which the parties can choose among \mbox{$d+1$} measurement bases. Within this framework, we consider: a \textit{\mbox{$d+1$}-basis spot-checking} protocol (denoted Protocol~\ref{protIb}), where one preferred basis is chosen with higher probability and used for the key generation rounds, while data from all the \mbox{$d+1$} bases is used for parameter estimation; and a \textit{\mbox{$d+1$}-basis with multiple key-generation basis} protocol (denoted Protocol~\ref{protIc}), where all bases contribute to both parameter estimation and key generation.

The key rate simulations consider an implementation where the distributed state is a $d$-dimensional maximally entangled state subjected to depolarizing noise with visibility $\nu$, and Alice’s detection efficiency is $\eta$.

%%%%%%%%%%%%%%%%%%%%%%%%%%%%%%%%%%%%%%%%%%%%%%%%%%%%%%%%%%%%%%%%%%%%%%%%%%%%%%%%%%%%%%%%%%%%%%%%
\subsubsection{Reverse reconciliation} \label{Reverse vs direct reconciliation}
\begin{figure}
\centering
\includegraphics[scale=0.475, keepaspectratio]{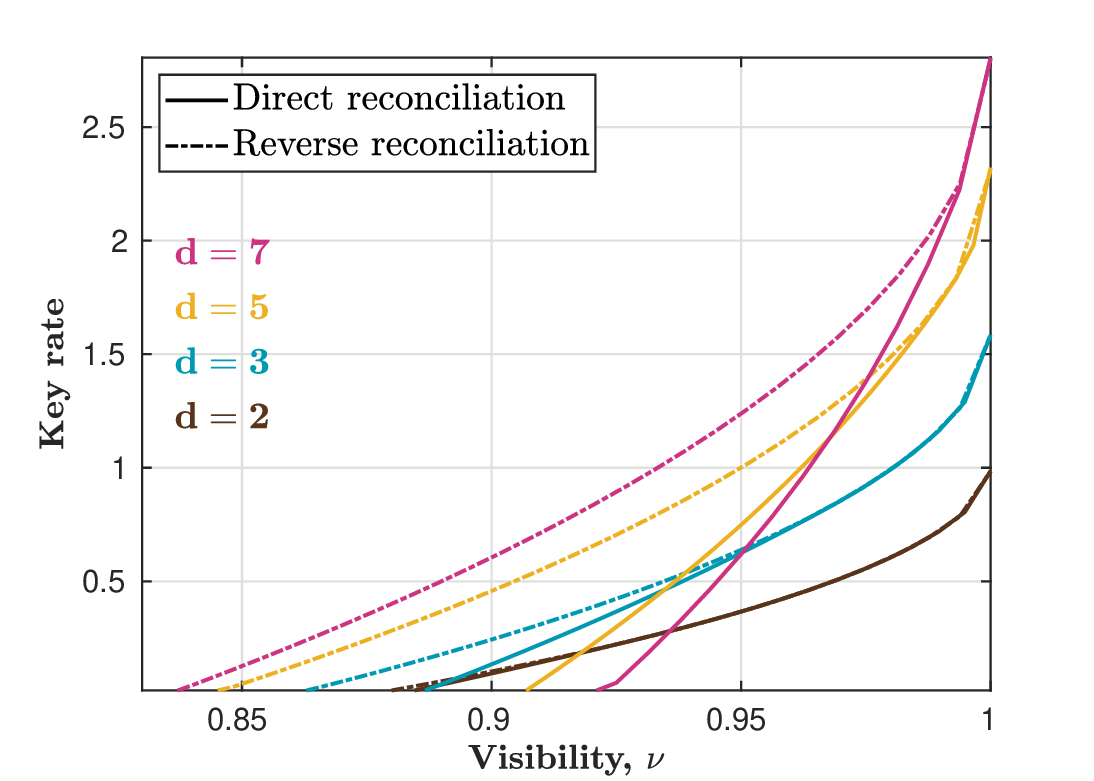}
\caption{{\bf Direct versus reverse reconciliation key rates as a function of visibility.}
Colors correspond to dimensions $d=2$ ({\bf\color{brown}{brown}}), $d=3$ ({\bf{\color{cyan}cyan}}), $d=5$ ({\bf{\color{yellow}yellow}}), and $d=7$ ({\bf{\color{magenta}magenta}}). Solid and dash-dot lines correspond to the direct reconciliation and reverse reconciliation key rates, respectively, for the spot-checking protocol with $m=d+1$ measurement settings.}
\label{infrecon_comparison}
\end{figure}

In QKD, one-way information reconciliation can be performed in two ways: \textit{direct reconciliation}, where Alice's string is taken as a reference, and Bob corrects his key accordingly, and \textit{reverse reconciliation}, where Bob's string is the reference, and Alice corrects her key accordingly. Now, unlike symmetric entanglement-based device-dependent and device-independent QKD protocols, where the key rate remains invariant under the choice of these two reconciliation strategies, 1sDI-QKD based on quantum steering exhibits inherent asymmetry between Alice and Bob. In fact, for the two-basis protocol, an analysis based on the EUR is only possible for the reverse reconciliation scenario, since we make use of the knowledge of Bob's device to estimate the overlap between the POVM elements corresponding to his two basis choices. 
Motivated by this asymmetry, we further investigated whether it can also be leveraged for \mbox{$d+1$}-basis protocols within the min-entropy framework. Our analysis reveals that reverse reconciliation indeed yields higher secret key rates than direct reconciliation, with the advantage becoming more pronounced in higher dimensions. This is illustrated in Fig.~\ref{infrecon_comparison}, which shows that for direct reconciliation, high-dimensional advantage is quickly lost with noise, while for reverse reconciliation, higher dimensions lead to higher key rates and greater noise tolerance.

Interestingly, when examining the guessing probabilities $P_{\mathrm{guess}}(A|E,X,Y)$ and $P_{\mathrm{guess}}(B|E,X,Y)$, we observe that $P_{\mathrm{guess}}(B|E,X,Y) < 1$ already occurs for  $\beta_{\mathrm{obs}} < \beta_{\mathrm{LHS}}$. On the other hand,  $P_{\mathrm{guess}}(A|E,X,Y) < 1$ only occurs for values of $\beta_{\mathrm{obs}}$ that certify steering. In Appendix~\ref{steering_detection_wrt_direct_reverse}, we investigate this phenomenon in more detail and show, for example, in the case of $d=2$ that the LHS bound for the steering inequality is $\beta_{\mathrm{LHS}}\approx1.71$. In the reverse-reconciliation scenario, however, $P_{\mathrm{guess}}(B|E,X,Y) < 1$ already occurs at $\beta_{\mathrm{obs}} \approx 1.51$, whereas in the direct-reconciliation scenario the corresponding threshold appears at the larger value $\beta_{\mathrm{obs}} \approx 1.71$. This indicates that randomness can be witnessed even before the steering inequality under consideration is violated.
Although this may seem counterintuitive at first, we remark that the secret key rate, and not the randomness, is a witness of entanglement \cite{entprecond_PhysRevLett.92.217903}. Consistently, we observe a positive key rate only once the steering inequality is violated.
Interestingly, we further analyze the properties of the optimal quantum states responsible for exhibiting randomness prior to violation of the steering inequality, and find that these states are in fact steerable. This point is discussed in detail in Appendix~\ref{steering_detection_wrt_direct_reverse}.

To quantify the contribution of error correction in a reverse reconciliation scheme, we consider the depolarizing-loss model where the distributed state is a noisy $d$-dimensional maximally entangled state with visibility $\nu$, and Alice’s detection efficiency is $\eta$. The associated leakage term, quantified by the conditional entropy $H(B|A,X,Y)$, is given by
\begin{equation}
    \begin{aligned}
        \rm{leak_{IR}^{\leftarrow}} & = -\eta\left(\nu + \frac{1-\nu}{d}\right)
                \log_2\!\left(\nu + \frac{1-\nu}{d}\right)\\
                 + (1-\eta)\log_2 & (d)  - \eta(d-1)\left(\frac{1-\nu}{d}\right)\log_2\!\left(\frac{1-\nu}{d}\right) \,.
    \end{aligned}
    \label{error_corr_term}
\end{equation}
For a detailed derivation, we refer to Appendix \ref{error_correction}.
The expression given in Eq.~(\ref{error_corr_term}) holds for all protocol variants considered in this work: the two-basis protocol (Protocol~\ref{protIa}), \mbox{$d+1$}-basis spot-checking protocol (Protocol~\ref{protIb}), and  \mbox{$d+1$}-basis with multiple key-generation basis protocol (Protocol~\ref{protIc}).

Finally, in Appendix~\ref{exp_inf_recon} we compare this model with experimental data and show that the conditional entropy $H(B|A, X ,Y)$ extracted from the observed statistics is in very good agreement with the predictions of the depolarizing-loss model given by Eq.~(\ref{error_corr_term}), with the two curves closely overlapping for all dimensions considered.

%%%%%%%%%%%%%%%%%%%%%%%%%%%%%%%%%%%%%%%%%%%%%%%%%%%%%%%%%%%%%%%%%%%%%%%%%%%%%%%%%%%%%%%%%%%%%%%%
\subsection{\textit{Two-basis protocol} (Protocol~\ref{protIa})} \label{section_twobasis_protocol}
In the two-basis variant of the protocol~\ref{box:protocol1}, Alice and Bob choose between two possible measurement bases in each round. They use the measurement setting $X=0$ and $Y=0$ for the key-generation basis, while $X=1$ and $ Y=1$ are used for parameter estimation. The key-generation basis is chosen with higher probability so that, in the asymptotic limit, it is used almost always, and the reduction in key rate due to sifting becomes negligible. The key rate is derived using the entropic uncertainty relation (EUR) approach introduced in \cite{Berta2010-ls}, which relates Eve's uncertainty about Bob's outcome in a certain basis ($Y=0$) with Alice's uncertainty about Bob's outcome in the other basis ($Y=1$):
\begin{equation}
    H(B|E,X=Y=0) \geq -\log_2 c - H(B|A,X=Y=1)\,,
    \label{eq:eur_relation}
\end{equation}
where $c$ quantifies the maximum overlap between Bob's measurement bases. For projective MUB measurements, $c = 1/d$, where $d$ denotes the dimension of the system. 

EUR has previously been applied to 1sDI-QKD in the qubit setting by Branciard \textit{et al.} \cite{branciard2012one}, where the secret key rate is evaluated on postselected data. We note, however, that postselection by an untrusted party cannot be performed without relying on further assumptions, since the information about which rounds were postselected could in principle be exploited by an adversary. We detail the problems with the approach of \cite{branciard2012one} in Appendix \ref{EUR_analytical}. Such an issue was also independently noted in \cite{lobo2026generalizedmeasurementincompatibility}.

The uncertainty relation, Eq.~\eqref{eq:eur_relation}, must therefore be applied to the full dataset. We extend the entropic security analysis to high-dimensional systems and adopt an extra-outcome strategy to model losses. Our analysis, therefore, does not rely on postselection. We will show that our strategy leads to significantly improved key rates and allows detection efficiencies as low as 50\% in the noiseless case, which corresponds to the minimum possible threshold for two measurement settings \cite{bennet_arbdeteff}. Substituting Eq.~(\ref{eq:eur_relation}) into the key rate expression (Eq.~(\ref{keyrateBE})), the asymptotic secret key rate of the protocol is given by
\begin{equation}
    \begin{aligned}
    r_\infty^{\mathrm{two\mbox{-}basis}} \geq& \log_2\!\left( d\right)- H(B|A,X=Y=1)\\
    &- H(B|A,X=Y=0)\,.\label{eq:keyrate_eur}
\end{aligned}
\end{equation}
where for a depolarizing-loss model we have that $H(B|A,X=Y=1)= H(B|A,X=Y=0)=  \rm{leak_{IR}^{\leftarrow}}$. The condition $\log_2\!\left( d\right)-H(B|A,\mbox{$X=Y=1$})-H(B|A,\mbox{$X=Y=0$})$ $\textgreater 0$ can be interpreted as a steering witness, since we are in a 1sDI setting. Therefore, a positive secret key rate implies entanglement of the shared state \cite{entprecond_PhysRevLett.92.217903}. 
%%%%%%%%%%%%%%%%%%%%%%%%%%%%%%%%%%%%%%%%%%%%%%%%%%%%%%%%%%%%%%%%%%%%%%%%%%%%%%%%%%%%%%%%%%%%%%%%
\subsection{\itshape{\mbox{$d+1$}-basis} protocols} \label{section_d+1basis_protocol}
Next, we consider protocols in which Alice and Bob can choose among \mbox{$d+1$} possible measurement bases. In this case, the security analysis is carried out using a min-entropy framework formulated as a semidefinite program (SDP) constrained by steering inequalities, which allows multiple measurement settings to be incorporated into the analysis. The min-entropy $H_{\mathrm{min}}(B|E,X,Y)$ provides a lower bound on the conditional von Neumann entropy $H(B|E,X,Y)$ and is directly related to Eve's guessing probability as 
\begin{equation}
    H_{\min}(B|E,X,Y) = -\log_2 P_{\mathrm {guess}}(B|E,X,Y)\,,
    \label{eq:Hmin_BE}
\end{equation}
where $P_{\mathrm {guess}}(B|E,X,Y)$ denotes the probability with which Eve can correctly guess Bob's outcomes. Here, we analyze the following two protocols:\\

\textit{(i) Spot-checking protocol} (Protocol~\ref{protIb}): In this protocol, a fixed measurement basis setting \mbox{$Y=y^*$} (and corresponding \mbox{$X=x^*$}) is chosen with higher probability and is used for key generation, while the statistics of all \mbox{$d+1$} bases are used for parameter estimation (i.e., to test for the violation of a steering inequality). To compute $P_{\mathrm{guess}}(B|E,X=x^*,Y=y^*)$, we express it as a function of the assemblage prepared after Alice’s and Eve’s measurements, denoted by $\sigma_{a|x}^{e}$, and optimize it via the SDP given below that incorporates the observed steering violation as a constraint (details provided in Appendix~\ref{fixed_basis}):
\begin{subequations}\label{eq:sdp_spotchecking}
\begin{align}
        \max_{\{\sigma_{a|x}^{e}\}} \quad & \sum_{a=0}^{d}\sum_{b=0}^{d-1} \mathrm{tr} \left[ N_{b|y^*} \, \sigma_{a|x^*}^{e=b} \right] \label{obfunc}\\
        \text{subject to} \quad
        &\sum_{a,x} \mathrm{tr} 
        \left[ N_{b=a|y=x} \sum_e \sigma_{a|x}^{e} \right] = \beta_{\mathrm{obs}} \label{ineq}\\
        &\sum_a \sigma_{a|x}^{e} = \sum_a \sigma_{a|x'}^{e} \quad \forall\, x \ne x',\, e \label{nosig}\\         
        &\sum_{a,e} \mathrm{tr} \left[\sigma_{a|x}^{e}\right] = 1 \quad \forall\, x \label{norm}\\
        &\sigma_{a|x}^{e} \geq 0 \quad \forall\, a, x, e \label{pos}\,,
\end{align}
\end{subequations}
where $N_{b|y}$,  for $b\in\{0,1,\ldots, d-1\}$, are Bob's measurement operators for measurement setting $Y=y$, here we consider them to be projective MUBs \cite{WOOTTERS1989363}, and $N_{\varnothing|y}=\frac{1}{\sqrt{d}}\mathds{I} $, see Eq.~\eqref{SIloss}. Eq.~(\ref{ineq}) ensures that the average assemblage prepared by Eve remains consistent with the observed violation $\beta_{\mathrm{obs}}$ of the steering inequality. Eq.~(\ref{nosig}) imposes no-signalling, i.e., Alice cannot signal to Bob and Eve. Eq.~(\ref{norm}) and Eq.~(\ref{pos}) impose normalization and positivity. Let $h_1$ denote the optimal value of the above SDP. Substituting $H_{\min}(B|E,\mbox{$Y=y^*$}) = -\log_2 h_1$ into Eq.~(\ref{keyrateBE}), and using the estimated leakage for the depolarizing-loss model, Eq.~\eqref{error_corr_term}, leads to the following lower bound on the key rate:
\begin{equation}
    \begin{aligned}
        r_\infty^{\mathrm{spot\mbox{-}checking}} \geq & -\log_2 (h_1) - \rm{leak_{IR}^{\leftarrow}}\,.
    \label{eq:keyrate_spotchecking}
    \end{aligned}
\end{equation}
\textit{(ii) Multiple key-generation basis protocol} (Protocol~\ref{protIc}): In this protocol, all $d+1$ measurement bases are used for both key generation and parameter estimation, chosen according to a distribution $p(x,y)$. Here we choose a uniform distribution of all the inputs, i.e., $p(x,y)=1/m$ for all $x,y$ $\in$ $\{0,\ldots,\mbox{$m-1$}\}$. The guessing probability averaged over all measurement settings of Bob is then given by
\begin{equation}
\begin{aligned}
    P_{\mathrm {guess}}(B|E,X,Y) &= \max_{\{E\}_y} \sum_{y\in Y} p(y)\quad  \\ \sum_{b} &P_{\mathrm{guess}}(B=E=b|Y=y)\,.
\end{aligned}
\end{equation}
The corresponding optimization problem can be written as the following SDP (details provided in Appendix~\ref{all_bases}):
\begin{subequations}\label{eq:sdp_multikeybases}
\begin{align}
        \max_{\{\sigma_{a|x}^{e,z}\}} \quad &\sum_{y\in Y} p(y) \sum_{a=0}^{d}\sum_{b=0}^{d-1} \mathrm{tr} \left[ N_{b|y} \, \sigma_{a|x^*}^{e=b,z=y} \right]\\
        \text{subject to} \quad
        &\sum_{a,x} \mathrm{tr} 
        \left[ N_{b=a|y=x} \sigma_{a|x} \right] = \beta_{\mathrm{obs}} \\
        &\sum_{e} \sigma_{a|x}^{e,z} = \sigma_{a|x} \quad \forall a,x,z \\
        &\sum_a \sigma_{a|x}^{e,z} = \sum_a \sigma_{a|x'}^{e,z} \quad \forall\, x \ne x',\, e, z\\         
        &\sum_{a,e,z} \mathrm{tr} \left[\sigma_{a|x}^{e,z}\right] = 1 \quad \forall\, x, z\\
        &\sigma_{a|x}^{e,z} \geq 0 \quad \forall\, a, x, e, z \,.
\end{align}
\end{subequations}
The constraints have analogous interpretations to those in Eq.~(\ref{eq:sdp_spotchecking}) (see Appendix~\ref{all_bases}). Let $h_2$ denote the optimal value of the above SDP, resulting in a key-rate expression
\begin{equation}
    \begin{aligned}        r_\infty^{\mathrm{multi\mbox{-}key\mbox{-}basis}} \geq & -\log_2 (h_2) - \rm{leak_{IR}^{\leftarrow}}\,,
    \label{eq:keyrate_multikeybasis}
    \end{aligned}
\end{equation}
where $\rm{leak_{IR}^{\leftarrow}}$ denotes the information reconciliation leakage for the depolarizing-loss model, given by Eq.~\eqref{error_corr_term}. Note that, for the symmetric noise model considered, the leakage is the same for all the considered protocols.

% SECTION IV %%%%%%%%%%%%%%%%%%%%%%%%%%%%%%%%%%%%%%%%%%%%%%%%%%%%%%%%%%%%%%%%%%%%%%%%%%%%%%%%%%
\section{High-dimensional advantage}
In this section, we evaluate the protocol's performance following the methodology outlined in Section~\ref{III}. Recognizing that ideal conditions cannot be achieved in practice, it is essential to account for the impact of noise and loss on the protocol's performance.
\vspace{1em}

\textit{Noise modelling:}
To model noise, we assume that Alice and Bob share an isotropic state subjected to depolarizing noise of the form $\nu |\phi^+_d\rangle\langle\phi^+_d| + (1-\nu)\mathds{I}/d^2 $, where $|\phi^+_d\rangle = \sum_{i=0}^{d-1}|ii \rangle/\sqrt{d}$ denotes the maximally entangled state in a $d$-dimensional Hilbert space. The parameter $\nu \in [0,1]$ represents the visibility of the ideal maximally entangled state, while $1-\nu$ quantifies the noise contribution. For the implementation of the QKD protocol, Alice and Bob perform projective MUB measurements $M_{a|x}=|\psi_a^x\rangle\langle\psi_a^x|$ and $N_{b|y}=|\psi_b^y\rangle\langle\psi_b^y|$ where $|\psi_i^j\rangle$ are MUB vectors defined in Ref. \cite{WOOTTERS1989363}, which for prime $d$ take the form,
\begin{equation}
|\psi_i^j\rangle=\frac{1}{\sqrt{d}} \sum_{l=0}^{d-1} \omega^{i l+j l^2}|l\rangle\,,
\end{equation}
with \(i\in\{a,b\}\) and \(j\in\{x,y\}\) and $\omega = \mathrm{exp}(2\pi i/d)$ the $d$th root of unity. For prime powers, a similar form holds, for which we use a construction provided in Ref.\cite{designolleMUB}. The first basis corresponds to the computational basis, $\{|l\rangle\}^{d-1}_{l=0}$, while the remaining $d$ bases are given by ${\{|\psi_i^j\rangle}\}^{d-1}_{i=0}$, with $j=1,\dots,d$. We remark that although these are the measurements we expect Alice to perform in the actual experimental implementation, they are not assumed in the security proof, as Alice’s device remains uncharacterized in the key rate optimization.
\vspace{1em}

\textit{Loss modelling:} Practical detectors have imperfections and may fail, resulting in no-click (no-detection) events. Additionally, channel losses may prevent photons from reaching the measurement devices. We denote by $\eta$ Alice's overall detection efficiency, incorporating both detection inefficiencies and channel losses. Since Alice's (untrusted) no-click events must be explicitly included in the analysis, we must choose a strategy to properly handle these events. This can be done using one of the following standard strategies:
i) \textit{random assignment strategy}: the no-click event $\varnothing$ is randomly assigned to one of the existing outcomes $\{0,1,\ldots,d-1\}$;
ii) \textit{deterministic assignment strategy}: the no-click event is mapped to a fixed outcome (e.g., $\varnothing = 0$); or
iii) \textit{extra-outcome strategy}: the no-click event is treated as an additional outcome, $\varnothing = \perp$, leading to \mbox{$d+1$} total outcomes.

We investigate the three approaches for handling no-click events and observe that the extra-outcome strategy consistently provides the best performance (see Appendix~\ref{analytical_bound_derivation} for a detailed comparison). Therefore, in the remainder of our analysis, we adopt the extra-outcome strategy. 

We use the loss-tolerant steering inequality~ \cite{loss-tolerant_SI_Paul} defined by the operators given in Eq.~\eqref{SIloss} (see Appendix \ref{analytical_bound_derivation} for details). The corresponding measurement operators on Alice's side, including no-click events as an extra outcome, can be modeled as
\begin{equation}
    M_{a|x}^{(\eta)} = 
    \begin{cases}
        \eta \, M_{a|x} & \text{for } a = 0, \dots, d-1 \\
        (1 - \eta) \, \mathds{I} & \text{for } a = \varnothing \,,
    \end{cases}
\end{equation}
where  $M_{a|x}$ are the projective MUB measurement operators. For the depolarizing-loss model considered here, characterized by visibility $\nu$ and detection efficiency $\eta$, the observed value of the steering functional is given by (refer to Appendix \ref{analytical_bound_derivation})
\begin{equation}
    \beta_{\mathrm{obs}}(\nu,\eta) = m \eta\left(\nu+\frac{1-\nu}{d}\right)+ \frac{m(1-\eta)}{\sqrt{d}}\,,
    \label{beta_{obs}}
\end{equation}
where $d$ denotes the system dimension and $m$ the number of projective MUB measurements. For the analysis of Protocols~\ref{protIb} and ~\ref{protIc}, we consider the complete set of MUBs, i.e., $m = d + 1$ for prime and prime-power dimensions. 

%%%%%%%%%%%%%%%%%%%%%%%%%%%%%%%%%%%%%%%%%%%%%%%%%%%%%%%%%%%%%%%%%%%%%%%%%%%%%%%%%%%%%%%%%%%%%%%%%%%%%
\subsection{Key Rate Analysis}
\subsubsection{Noise and loss thresholds}

\begin{figure}[h]
\centering
\includegraphics[scale=0.475, keepaspectratio]{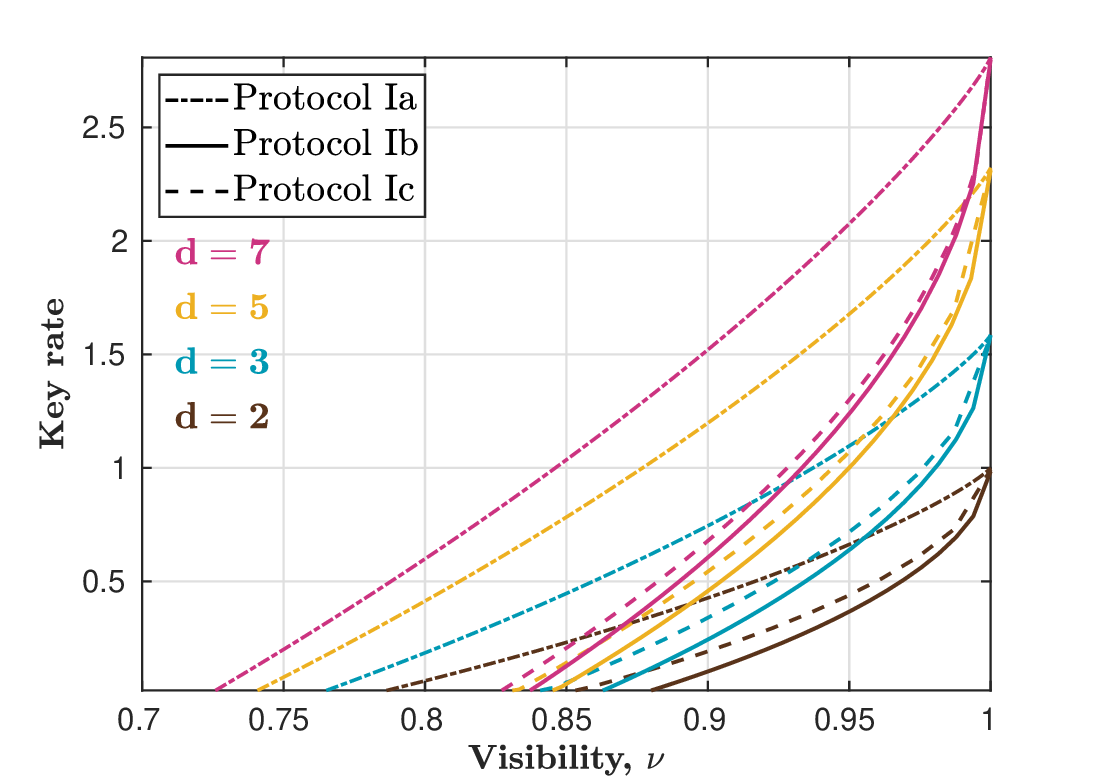}
\caption{{\bf Comparison of key rates as a function of visibility $\nu$ under depolarizing noise, assuming ideal detection efficiency ($\eta=1$).} Colors correspond to dimensions $d=2$ ({\bf\color{brown}{brown}}), $d=3$ ({\bf{\color{cyan}cyan}}), $d=5$ ({\bf{\color{yellow}yellow}}), and $d=7$ ({\bf{\color{magenta}magenta}}). Dash-dot, solid, and dashed lines correspond to Protocols~\ref{protIa} (two-basis), \ref{protIb} (spot-checking), and \ref{protIc} (multi-key-basis), respectively.}
\label{keyrate_noise_comparison}
\end{figure}

\begin{figure}
\centering
\includegraphics[scale=0.475, keepaspectratio]{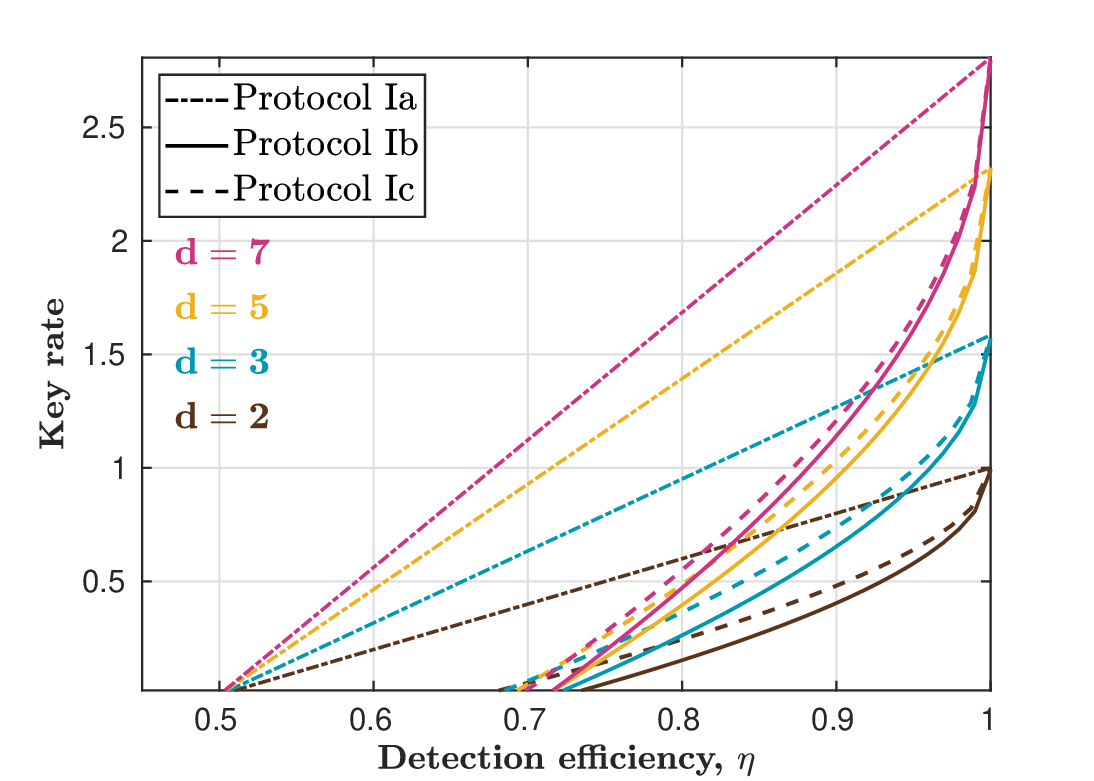}
\caption{{\bf Comparison of key rates as a function of detection efficiency $\eta$ under extra outcome strategy, assuming ideal visibility ($\nu=1$).} Colors correspond to dimensions ({\bf\color{brown}{brown}}), $d=3$ ({\bf{\color{cyan}cyan}}), $d=5$ ({\bf{\color{yellow}yellow}}), and $d=7$ ({\bf{\color{magenta}magenta}}). Dash-dot, solid, and dashed lines correspond to Protocols~\ref{protIa} (two-basis), \ref{protIb} (spot-checking), and \ref{protIc} (multi-key-basis), respectively.}
\label{keyrate_loss_comparison}
\end{figure}

In Fig.~\ref{keyrate_noise_comparison} and Fig.~\ref{keyrate_loss_comparison}, we compare the secret key rates as a function of visibility (for $\eta=1$) and detection efficiency  (for $\nu=1$), respectively, for all three protocols. The dash-dot lines correspond to the key rates obtained from Eq.~(\ref{eq:keyrate_eur}) for Protocol~\ref{protIa} (two-basis protocol). The solid lines correspond to the key rates obtained from Eq.~(\ref{eq:keyrate_spotchecking}) for Protocol~\ref{protIb} (\mbox{$d+1$}-basis spot-checking protocol), and the dashed lines represent the key rates obtained from Eq.~(\ref{eq:keyrate_multikeybasis}) for Protocol~\ref{protIc} (\mbox{$d+1$}-basis multi-key-basis protocol).

Fig.~\ref{keyrate_noise_comparison} shows that, for all three protocols, increasing the system dimension ($d$) leads to a significant enhancement in the maximum tolerable noise, quantified by the critical visibility $\nu_{\mathrm{cr}}$. Moreover, in the high-visibility regime, the key rate increases rapidly with dimension and approaches its theoretical maximum of $\log_2(d)$ bits in the limit of perfect visibility.
Fig.~\ref{keyrate_loss_comparison} illustrates the dependence of the key rate on detection efficiency. For Protocol~\ref{protIa}, the minimum required detection efficiency converges to the same value for all dimensions. This threshold can be derived analytically from Eq.~\eqref{eq:keyrate_eur}, which gives
\begin{equation}
    \eta_{\mathrm{cr}}^{\mathrm{(two\mbox{-}basis)}} > \frac{\log_2 d}{2\left[\log_2 d+a \log_2 a+(d-1)b \log_2 b\right]}\,,
    \label{eta_cr_EUR}
\end{equation}
where $a = \nu + \frac{1-\nu}{d}$ and $b = \frac{1-\nu}{d}$. For $\nu=1$, this simplifies to $\eta_{\mathrm{cr}}^{\mathrm{(two\mbox{-}basis)}} \textgreater 1/2$ for all values of $d$. This bound is in fact tight, as it has been shown in~\cite{Acin2016necessarydetectionefficiencies} that spot-checking protocols become insecure for detection efficiencies $\eta \leq 1/2$, independently of the system dimension and the number of measurement settings.  

For Protocol~\ref{protIb} (solid curves), the loss tolerance, quantified by minimum detection efficiency $\eta_{cr}$, improves slightly with increasing dimension (see Fig.~\ref{keyrate_loss_comparison}). In contrast, for Protocol~\ref{protIc} (dashed curves), the loss tolerance is higher than for Protocol~\ref{protIb}, but slightly degrades as the dimension increases (see Fig.~\ref{keyrate_loss_comparison}). However, as we show in Section \ref{trade-off_section}, in the presence of noise, the critical detection efficiency improves with increasing dimension.
Overall, the EUR-based analysis for the two-basis protocol leads to the highest key rates, owing to the tight lower bounds it provides on the conditional von Neumann entropy. In contrast, our analysis for the \mbox{$d+1$}-basis protocols relies on min-entropy bounds, which are generally not tight. However, unlike the EUR framework, which is inherently restricted to two measurement bases, the min-entropy approach can flexibly accommodate different settings and therefore enables the analysis of Protocols~\ref{protIb} and~\ref{protIc}, which employ \mbox{$d+1$} measurement bases.

\begin{figure*}
\centering
\begin{subfigure}[t]{0.48\textwidth}
    \includegraphics[width=\linewidth, keepaspectratio]{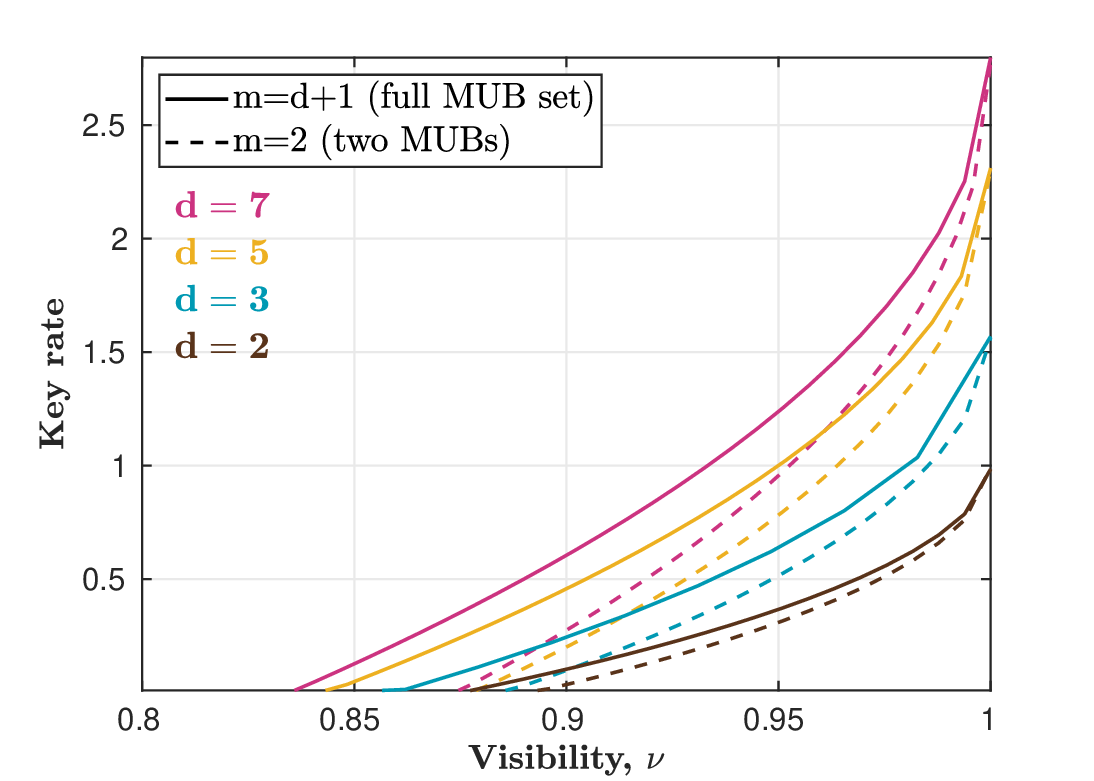}
    \caption{}\label{vissubplot_multi-measuremnt_advantage}
\end{subfigure}
\hfill
\begin{subfigure}[t]{0.48\textwidth}
    \includegraphics[width=\linewidth, keepaspectratio]{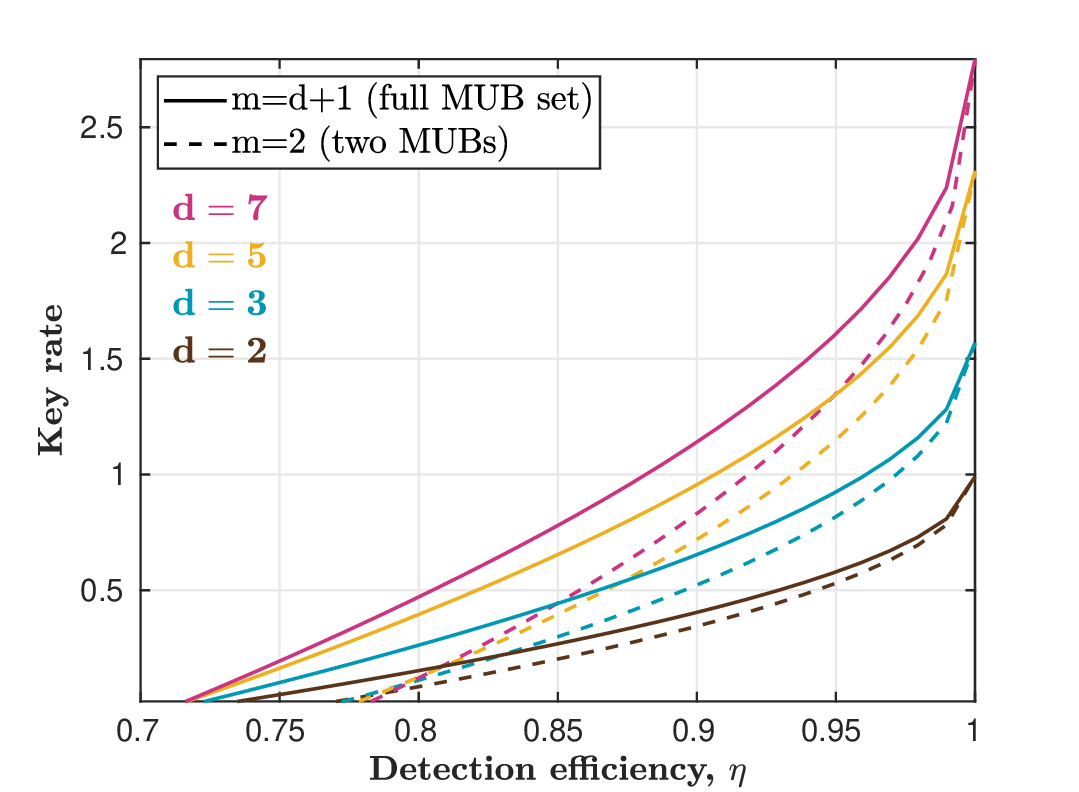}
    \caption{} \label{effsubplot_multi-measuremnt_advantage}
\end{subfigure}
\caption{Key-rate advantage of using the full set of MUBs ($m=d+1$) over two MUBs ($m=2$) for \mbox{$d+1$}-basis spot-checking protocol (\ref{protIb}). (a) Key rate as a function of visibility $\nu$ under depolarizing noise, assuming ideal detection efficiency ($\eta=1$). (b) Key rate as a function of detection efficiency $\eta$, assuming ideal visibility ($\nu=1$). Colors correspond to dimensions $d=2$ ({\bf\color{brown}{brown}}), $d=3$ ({\bf{\color{cyan}cyan}}), $d=5$ ({\bf{\color{yellow}yellow}}), and $d=7$ ({\bf{\color{magenta}magenta}}). Solid lines correspond to steering correlations using $m=d+1$ MUB measurement settings in dimension $d$, while dashed lines correspond to correlations using two MUB measurement settings ($m=2$).}
\label{multi_measurement_advantage}
\end{figure*}

Although the resulting min-entropy bounds are not sufficient to outperform the EUR-based key rates, they can provide insight into the expected behavior of the protocols. Fig.~\ref{multi_measurement_advantage} compares the key rates of the two-basis protocol (Protocol~\ref{protIa}) and the \mbox{(d+1)}-basis spot-checking protocol (Protocol~\ref{protIb}) when the same min-entropy framework is applied to analyze both protocols. We see that, for all ranges of parameters, the $d+1$-basis protocol exhibits significantly improved performance. This demonstrates a clear advantage of using multiple measurement bases at the level of min-entropy bounds, suggesting that a similar improvement may persist for the optimal key rates.

Additional evidence supporting the potential of  $d+1$-basis protocols comes from the behaviour of steering thresholds. Increasing the number of measurement settings generally lowers the requirements for demonstrating steering. For the loss-tolerant steering inequality considered here, the required detection efficiency for steering violation is given by~\cite{loss-tolerant_SI_Paul}:
\begin{equation}
    \eta_{\mathrm{req}}^{\mathrm{(d+1\mbox{-}basis)}} \textgreater \frac{\sqrt{d}}{m\left(\nu\left(\sqrt{d}+1\right)-1\right)}\,.
    \label{eta_cr_SI}
\end{equation}
In the noiseless case ($\nu=1$), this simplifies to $\eta^{\mathrm{(d+1\mbox{-}basis)}}_{\mathrm{req}} \textgreater 1/m$.
Similarly, the visibility required to demonstrate steering is
\begin{equation}
    \nu_{\mathrm{req}}^{\mathrm{(d+1\mbox{-}basis)}} \textgreater \frac{1+(\sqrt{d}/m\eta)}{\sqrt{d}+1} \,.
    \label{nu_cr_SI}
\end{equation}
These thresholds characterize only the emergence of steering and do not, in general, imply positive secret key rates. Nevertheless, they suggest that increasing the number of measurement settings can relax the minimum requirements on both detection efficiency and visibility needed to demonstrate the correlations relevant for 1sDI-QKD, thereby indicating the potential for more robust key rates. This observation is particularly interesting in light of known attacks~\cite{Acin2016necessarydetectionefficiencies}: while spot-checking protocols are fundamentally limited by attacks for detection efficiency below $\eta\leq 1/2$, protocols employing multiple key-generating bases are known to admit attacks only up to $\eta\leq 1/m$. Therefore, Protocol~\ref{protIc} is not constrained by the same $50\%$ threshold and may in principle achieve positive key rates below $\eta=1/2$.

We conjecture that Protocols~\ref{protIb} and~\ref{protIc} can indeed outperform the two-basis protocol. However, establishing this rigorously would require tighter key rate bounds for the $d+1$-basis protocols. Since no efficient general techniques are currently known to tightly bound the conditional von Neumann entropy in the 1sDI setting for higher dimensions, this remains an open problem for future work. Recent numerical techniques \cite{Brown2021-pa, Navascues_2008} provide promising routes for extending SDP-based analyses to obtain tighter bounds on the conditional von Neumann entropy.

\subsubsection{Noise-loss trade-off in higher dimensions}\label{trade-off_section}
In a practical setting, both noise and loss are present simultaneously, and thus it is important to analyze how these two parameters trade off against each other as the system dimension varies. Table~\ref{tab:vis_cr_table} \&~\ref{tab:eta_cr_table} show the dependence of the critical visibility $\nu_{\mathrm{cr}}$ and critical detection efficiency $\eta_{\mathrm{cr}}$ on the system dimension for fixed values of detection efficiency and visibility, respectively, for Protocols~\ref{protIa}, \ref{protIb}, and \ref{protIc}.

\begin{table}
\centering
\caption{Dimensional scaling of critical visibility $\nu_{\mathrm{cr}}$ thresholds for protocols~\ref{protIa} (two-basis), \ref{protIb} (spot-checking), and \ref{protIc} (multi-key-basis) against fixed detection efficiencies denoted by $\eta$.}
\label{tab:vis_cr_table}
\centering
\begin{tabular}{
c
>{\centering\arraybackslash}p{0.4cm}
>{\centering\arraybackslash}p{1.5cm}
>{\centering\arraybackslash}p{1.5cm}
>{\centering\arraybackslash}p{1.5cm}
}
\hline
\rule{0pt}{3ex}
\textbf{Critical visibility} & \textbf{$d$}
& $\eta=1.0$
& $\eta=0.90$
& $\eta=0.80$ \\
\hline
\rule{0pt}{2.8ex}
\multirow{4}{*}{\begin{tabular}{c}
$\nu_{\mathrm{cr}}^{\mathrm{two\mbox{-}basis}}$\\{\footnotesize (Protocol~\ref{protIa})}\\
\end{tabular}}
& 2 & 0.779 & 0.815 & 0.855 \\
& 3 & 0.760 & 0.798 & 0.840 \\
& 5 & 0.737 & 0.778 & 0.823 \\
& 7 & 0.723 & 0.764 & 0.812 \\
& 9 & 0.712 & 0.755 & 0.804 \\
\hline
\rule{0pt}{2.8ex}
\multirow{4}{*}{\begin{tabular}{c}
$\nu_{\mathrm{cr}}^{\mathrm{spot\mbox{-}checking}}$\\ {\footnotesize (Protocol~\ref{protIb})}\\
\end{tabular}}
& 2 & 0.871 & 0.914 & 0.963\\
& 3 & 0.844 & 0.903 & 0.956\\
& 5 & 0.841 & 0.891 & 0.948\\
& 7 & 0.834 & 0.884 & 0.945\\
& 9 & 0.829 & 0.882 & 0.943 \\
\hline
\rule{0pt}{2.8ex}
\multirow{4}{*}{\begin{tabular}{c}
$\nu_{\mathrm{cr}}^{\mathrm{multi\mbox{-}key\mbox{-}basis}}$\\{\footnotesize (Protocol~\ref{protIc})}\\
\end{tabular}}
& 2 & 0.846 & 0.889 & 0.936 \\
& 3 & 0.833 & 0.883 & 0.933 \\
& 5 & 0.828 & 0.877 & 0.933 \\
& 7 & 0.825 & 0.875 & 0.933 \\
& 9 & 0.821 & 0.873 & 0.935 \\
\hline
\end{tabular}
\end{table}

As shown in Table~\ref{tab:vis_cr_table}, the critical visibility generally decreases with increasing dimension for all three protocols, demonstrating the overall high-dimensional advantage. A slight deviation from this trend is observed for $d=9$ at $\eta=0.80$ in Protocol~\ref{protIc}. This, however, may be attributable to numerical precision in the SDP optimization. 
Table~\ref{tab:eta_cr_table} shows the corresponding behavior of the critical detection efficiency as dimensions increase. For Protocol~\ref{protIa}, the threshold remains fixed at $\eta_{\mathrm{cr}}=0.5$ in the noiseless case ($\nu=1$), while for non-unit visibilities the required detection efficiency decreases with increasing dimension, highlighting high-dimensional advantage. 

\begin{table}
\centering
\caption{Dimensional scaling of critical detection efficiency $\eta_{\mathrm{cr}}$ thresholds for protocols~\ref{protIa} (two-basis), \ref{protIb} (spot-checking), and \ref{protIc} (multi-key-basis) against fixed visibilities denoted by $\nu$.}
\label{tab:eta_cr_table}
\centering
\begin{tabular}{
c
>{\centering\arraybackslash}p{0.4cm}
>{\centering\arraybackslash}p{1.5cm}
>{\centering\arraybackslash}p{1.5cm}
>{\centering\arraybackslash}p{1.5cm}
}
\hline
\rule{0pt}{3ex}
\textbf{Critical efficiency} & \textbf{$d$}
& $\nu=1.0$
& $\nu=0.95$
& $\nu=0.90$ \\
\hline
\rule{0pt}{2.8ex}
\multirow{4}{*}{\begin{tabular}{c}
$\eta_{\mathrm{cr}}^{\mathrm{two\mbox{-}basis}}$\\{\footnotesize (Protocol~\ref{protIa})}\\
\end{tabular}}
& 2 & 0.50 & 0.601 & 0.70 \\
& 3 & 0.50 & 0.591 & 0.68 \\
& 5 & 0.50 & 0.580 & 0.659 \\
& 7 & 0.50 & 0.575 & 0.648 \\
& 9 & 0.50 & 0.571 & 0.641 \\
\hline
\rule{0pt}{2.8ex}
\multirow{4}{*}{\begin{tabular}{c}
$\eta_\mathrm{cr}^{\mathrm{spot\mbox{-}checking}}$\\{\footnotesize (Protocol~\ref{protIb})}\\\end{tabular}}
& 2 & 0.723 & 0.827 & 0.938 \\
& 3 & 0.715 & 0.811 & 0.910 \\
& 5 & 0.712 & 0.797 & 0.886 \\
& 7 & 0.712 & 0.792 & 0.875 \\
& 9 & 0.713 & 0.791 & 0.871 \\
\hline
\rule{0pt}{2.8ex}
\multirow{4}{*}{\begin{tabular}{c}
$\eta_\mathrm{cr}^{\mathrm{multi\mbox{-}key\mbox{-}basis}}$\\{\footnotesize (Protocol~\ref{protIc})}\\
\end{tabular}}
& 2 & 0.669 & 0.771 & 0.875\\
& 3 & 0.677 & 0.771 & 0.866\\
& 5 & 0.687 & 0.773 & 0.859\\
& 7 & 0.694 & 0.775 & 0.856\\
& 9 & 0.700 & 0.777 & 0.856 \\

\hline
\end{tabular}
\end{table}
The minimum detection efficiency shows an overall decrease with increasing dimension for Protocol~\ref{protIb}, demonstrating the high-dimensional advantage in loss tolerance. The only deviation from this trend occurs for $d=9$ at $\nu = 1$ in Protocol~\ref{protIb}, analogous to the behavior observed in Table~\ref{tab:vis_cr_table} for Protocol~\ref{protIc}, and is likewise likely attributable to numerical precision in the SDP optimization.
In contrast, Protocol~\ref{protIc} shows a visibility-dependent dimensional advantage, with the critical detection efficiency increasing slightly with dimension for $\nu=1$ and $\nu=0.95$, while it decreases for the lower visibility $\nu=0.9$. Overall, these results demonstrate the robustness of high-dimensional 1sDI-QKD protocols in realistic scenarios where both noise and losses are unavoidable. The favorable scaling with dimension further supports the use of high-dimensional quantum systems as a promising approach for practical implementations.

\begin{figure*}[ht]
    \centering
    \includegraphics[width=\textwidth]{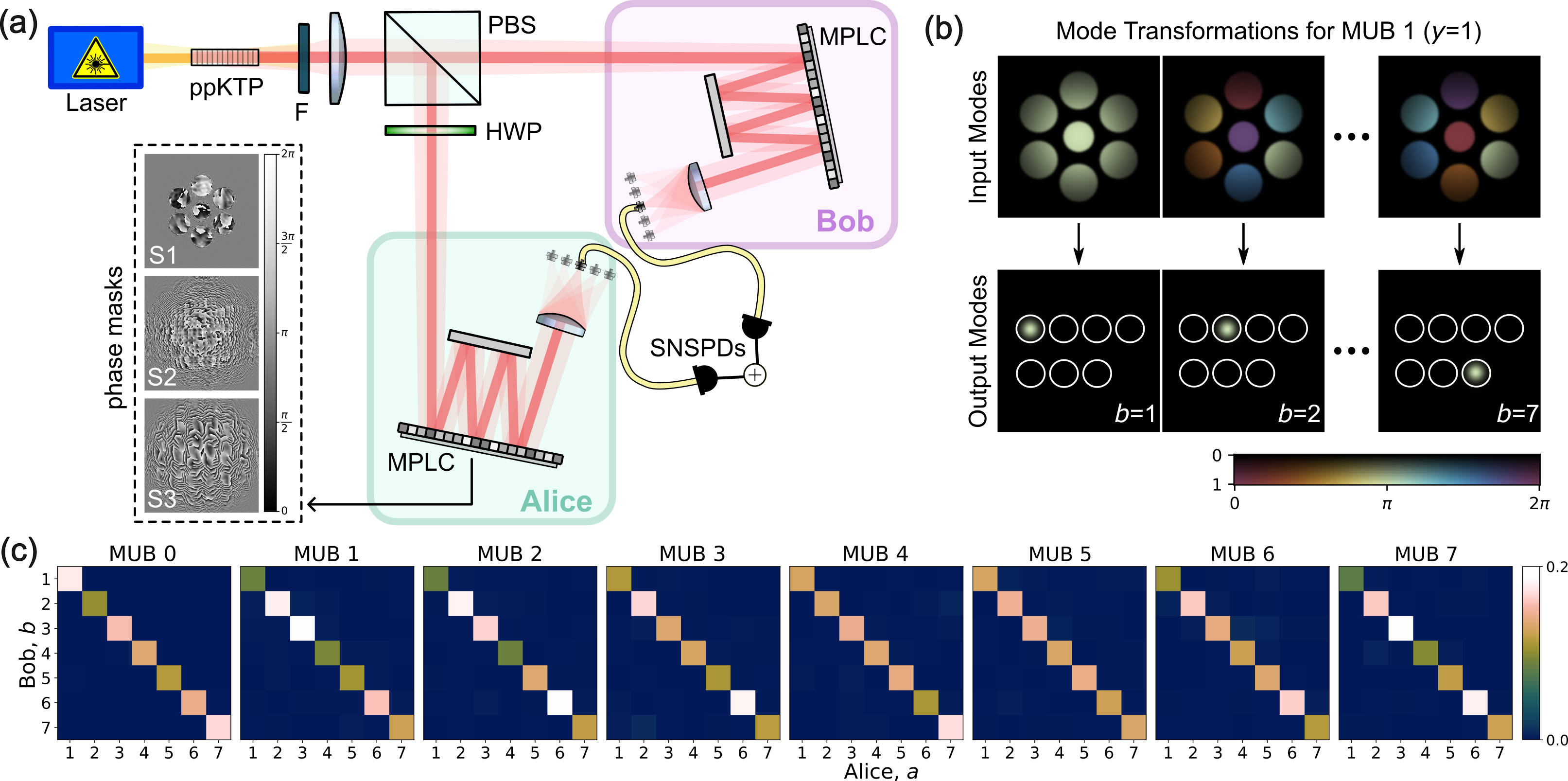}
    \caption{(a) Schematic representation of the experimental setup. A pair of spatially entangled photons at telecom wavelength~(1550 nm), generated by pumping a periodically poled potassium tri-phosphate (ppKTP) crystal, are distributed between Alice (cyan) and Bob (purple) after filtering out the pump. Alice and Bob are each equipped with a 3-plane multi-plane light converter (MPLC), which allows them to perform high-dimensional multi-outcome measurements across all MUBs. Coincidence counts between Alice and Bob are obtained via fiber-coupled SNSPD detectors. MPLC phase masks employed in Alice's MPLC (identical to those employed in Bob's MPLC) are shown as an inset. (b) A simulation of input and output modes at each MPLC. The first, second, and seventh input macro-pixel spatial modes for MUB 1 ($y=1$) in $d=7$ are each transformed into distinct output Gaussian spots arranged in a rectangular grid via the masks in (a), thus corresponding to distinct outputs. (c) Normalized two-photon coincidence counts showing correlations across all 8 MUBs in $d=7$ with an average visibility of $95.3\%$. Data for all other investigated dimensions are in presented in Fig.~\ref{all_mub_data}, Appendix \ref{corr_all_mub_data}.}
    \label{expt_setup_basic}
\end{figure*}

% SECTION V %%%%%%%%%%%%%%%%%%%%%%%%%%%%%%%%%%%%%%%%%%%%%%%%%%%%%%%%%%%%%%%%%%%%%%%%%%%%%%%%%%
\section{Experimental Implementation}

We demonstrate a proof-of-principle implementation of the high-dimensional 1sDI-QKD protocols analyzed in the previous sections using photons entangled in the transverse-spatial degree-of-freedom. The key parts of our demonstration include a high-quality source of spatial-mode entanglement and a programmable multi-outcome measurement, which is critical for any practical implementation of a high-dimensional QKD protocol. In our implementation, we use a mode conversion device known as a multi-plane light converter~(MPLC) to perform generalized multi-outcome measurements in the transverse-spatial degree-of-freedom~\cite{goel2023simultaneously,goel2026quantum}.

Figure ~\ref{expt_setup_basic}(a) shows a schematic representation of our experimental setup. The entangled pair of photons are generated at 1550nm via type-II spontaneous parametric down-conversion~(SPDC) by pumping a periodically-poled Potassium Titanyl Phosphate (ppKTP) crystal with a continuous wave pump laser centered at $775$ nm.
These photons are then distributed to two parties, Alice and Bob, who perform multi-outcome projective measurements across all MUBs in a chosen spatial-mode subspace.
In order to choose a suitable modal subspace, we characterize the joint-transverse momentum amplitude~(JTMA) of the biphoton state arising from the SPDC~\cite{srivastav2022characterizing}. 
Informed by the JTMA, we choose a set of spatially structured macro-pixel modes for dimensions $d = \{2,3,5,7,9,11\}$ to target maximal entanglement in the respective subspace~\cite{valencia2020high}.

The MPLCs are programmed to sort the bespoke set of macro-pixel spatial modes in a given MUB basis into a spatially-separated array of Gaussian spots, each corresponding to a measurement outcome~\cite{goel2023simultaneously} (see  Fig.~\ref{expt_setup_basic}(b) for an example). The joint-detection statistics are measured using a pair of detectors recording clicks for each pair of outcomes (see Appendix \ref{exp_app_details} for further details).

Each MPLC is implemented by placing a spatial light modulator~(SLM) parallel to a mirror such that all incoming photons are subjected to reflections from three consecutive phase-masks, each followed by a fixed distance of free-space propagation. 
The SLM allows us to reconfigure the phase masks, which redirect the light within the MPLC to perform a given spatial-mode transformation. 
We use an inverse-design algorithm called wavefront matching to calculate the phase masks corresponding to each measurement, respectively for each Alice and Bob~\cite{hashimoto_optical_2005, sakamaki_new_2007, goel2024inverse} (see Fig.~\ref{expt_setup_basic}(a) inset for an example). 
An example measurement for all MUBs in dimension $d=7$ is shown in Fig.~\ref{expt_setup_basic}(c), where the normalized correlation matrices quantify the probability distribution of two-photon coincidences between Alice and Bob for each measurement outcome. 

The observed coincidences, along with the exclusive single photons measured at Bob's detector, i.e., for which Alice has detected no photon and has thus recorded a no-click event, enable the steering inequality to be evaluated. Note that the no-click element of the steering inequality (identity matrix) can be realized by the completeness of any measurement at Bob (c.f. Eq. (\ref{SIloss})). After appropriate normalization by the total number of events, the observed steering violation $\beta_{\mathrm{obs}}$ can be calculated. See Appendix \ref{data_processing_details} for further details.

\begin{figure}[t]
\centering
\includegraphics[width=\linewidth]{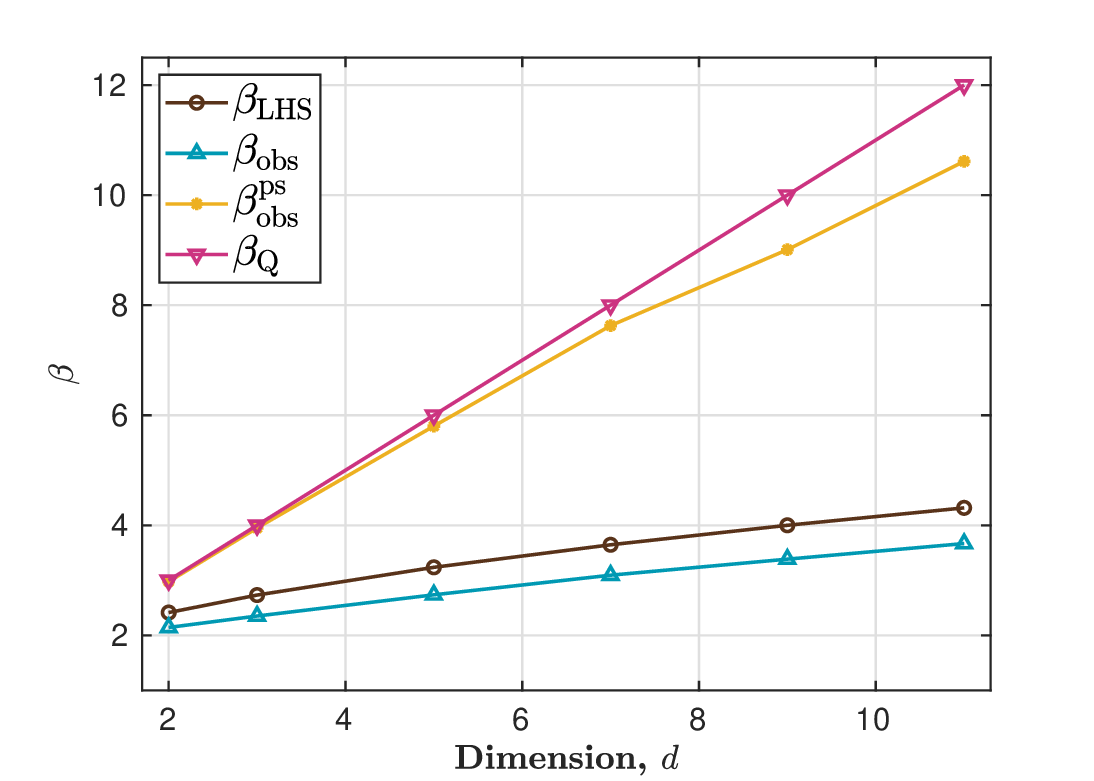}
\caption{Steering functional $\beta$ vs dimension $d$ up to $d=11$. Observed steering functional upon post-selection ({\bf\color{yellow}yellow} asterisks, $\beta^{\mathrm{ps}}_{\mathrm{obs}}$) violates the bound close to the quantum limit ({\bf\color{magenta}magenta} inverted triangles, $\beta_Q=d+1$). However, due to poor detection efficiencies owing to losses in the optical system, the observed steering functional from raw data ({\bf\color{cyan}cyan} triangles, $\beta_{\mathrm{obs}}$) does not violate the bound ({\bf\color{brown}brown} circles, $\beta_{\mathrm{LHS}}=1+\sqrt{d}$)}
\label{beta_vs_dim}
\end{figure}

If one is unconcerned with the opening of the detection loophole, i.e., under the fair-sampling assumption, we can consider the post-selected data, in which both photons are successfully detected. The corresponding observed post-selected steering functional, $\beta_{\mathrm{obs}}^{\mathrm{ps}}$, contains contributions only due to state and measurement imperfections. These data show significant violation of the steering inequality in all recorded dimensions (see  Fig.~\ref{beta_vs_dim}), and enable positive key rates for all considered dimensions (Tab.~\ref{table1}). Moreover, under the current noise levels, a clear high-dimensional advantage is observed up to dimension $d=7$.

Fig.~\ref{beta_vs_dim} also shows that, at the current loss level, a violation of the steering inequality can only be observed under the fair-sampling assumption.
Therefore, for a fully 1sDI demonstration, it is imperative to overcome device losses in future experimental implementations. In Appendix~\ref{losses_explained} we list the main instrumental efficiencies contributing to the overall system efficiency and discuss potential near-term improvements that could substantially reduce these losses.

% LOSS BUDGET

\begin{table}[t]
\centering
\begin{tabular}{
c
>{\centering\arraybackslash}p{1.5cm}
>{\centering\arraybackslash}p{1.9cm}
>{\centering\arraybackslash}p{1.9cm}
>{\centering\arraybackslash}p{1.9cm}
}
\toprule
$d$ & $\nu_\mathrm{exp}$ & $r_\infty^{\mathrm{two\mbox{-}basis}}$ & $r_\infty^{\mathrm{spot\mbox{-}checking}}$ & $r_\infty^{\mathrm{multi\mbox{-}key\mbox{-}basis}}$\\
\midrule
2  & 0.9846 & 0.8697 & 0.6538 & 0.7024\\ 
3  & 0.9817 & 1.3705 & 1.0211 & 1.0742\\ 
5  & 0.9589 & 1.7731 & 1.1241 & 1.1900\\ 
7  & 0.9458 & 2.0249 & 1.1760 & 1.2376\\ 
9  & 0.8884 & 1.6418 & 0.5808 & 0.6465\\ 
11 & 0.8729 & 1.6589 & 0.4773 & -\\ 
\bottomrule
\end{tabular}
\caption{Experimentally observed visibilities and asymptotic key rates calculated from the experimental data under fair-sampling assumption for Protocols~\ref{protIa} (two-basis), \ref{protIb} (spot-checking) and \ref{protIc} (multi-key-basis). Due to computational limitations, the key rate for $d=11$ for Protocol~\ref{protIc} could not be generated.}
\label{table1}
\end{table}

% SECTION VI %%%%%%%%%%%%%%%%%%%%%%%%%%%%%%%%%%%%%%%%%%%%%%%%%%%%%%%%%%%%%%%%%%%%%%%%%%%%%%%%%%
\section{Discussion}

We have presented a systematic security analysis of high-dimensional 1sDI-QKD protocols based on quantum steering and demonstrated a proof-of-principle experimental implementation using high-dimensional entanglement encoded in the transverse spatial degree-of-freedom of photon pairs and measured with a bespoke, programmable multi-outcome measurement device operating in up to dimension $d=11$. 

Our security analysis shows that high-dimensional encoding provides a promising route toward practical 1sDI-QKD by exploiting the enhanced robustness of high-dimensional entanglement against noise and loss. In particular, we show that reverse reconciliation yields substantially higher secret key rates than direct reconciliation in the 1sDI setting. Overall, our results demonstrate a clear high-dimensional advantage over a broad range of noise and loss regimes. The EUR-based analysis provides dimension-independent security down to a detection efficiency of $50\%$ by treating Alice's no-click events as an additional measurement outcome. This threshold coincides with the fundamental efficiency required for security in spot-checking protocols~\cite{Acin2016necessarydetectionefficiencies}.

Our min-entropy analysis further indicates that protocols employing multiple mutually unbiased bases can outperform two-basis protocols. Although the current min-entropy bounds for Protocols~\ref{protIb} and~\ref{protIc} do not yet surpass those obtained from the EUR approach, we conjecture that tighter secret key rate bounds for multi-basis protocols could reveal an even greater tolerance to noise and losses than is achievable with the EUR analysis for Protocol~\ref{protIa}. Establishing such bounds, therefore, constitutes an important direction for future work. More generally, deriving analytical bounds for the key rate would enable the treatment of arbitrarily large dimensions and provide a deeper understanding of the scaling of HD 1sDI-QKD.

Our proof-of-principle implementation demonstrates positive secret key rates for all three protocols in every investigated dimension up to $d=11$. We find that optimal rates are achieved in dimension $d=7$, which is directly related to the dimension-dependent quality of our multi-outcome measurements. As predicted by theory, this could be improved by optimizing the design of our measurement device via other algorithmic approaches such as gradient ascent or live optimization \cite{kupianskyi2023highdimensional,rocha2025selfconfiguring}.

Where losses are to be treated in their entirety, our current experimental implementation sits somewhat below the critical efficiency required for a loophole-free 1sDI QKD demonstration (See Fig.~\ref{beta_vs_dim} and Appendix \ref{simulating_eta}).
This is caused in part by the challenge of realizing  MPLCs with reconfigurable modulation devices~(i.e., SLMs).
Our implementation of the three-plane MPLC in dimension $d=7$ has an average efficiency of approximately 35\% (-4.56 dB loss), which is about seven times higher than recently achieved in multi-outcome measurements for device-dependent HD-QKD utilizing a ten-plane MPLC~\cite{Lib:25} (see Appendix \ref{losses_explained} for further details). 
This reduction in both optical complexity and loss is particularly important for 1sDI-QKD, where the key rate and the ability to demonstrate steering are highly sensitive to detection efficiency.
Other ways to further reduce losses include incorporating low-loss, static designs for MPLCs that use lithographically etched~\cite{fontaine2021hermite} or gold-plated  phase-planes~\cite{fang2021performance} and SNSPDs arrays~\cite{guardiani2024superconducting,fleming2025high}. 

Looking ahead, loophole-free demonstrations of steering violation, as well as finite-key analyses, will be essential for fully establishing the performance of HD 1sDI-QKD in realistic settings. Overall, our findings indicate that increasing the system dimension relaxes both the visibility and detection-efficiency requirements. The favorable dimensional scaling, together with the minimal assumptions on Alice’s devices, highlights high-dimensional 1sDI-QKD as a strong candidate for practical quantum communication.

% SECTION VII %%%%%%%%%%%%%%%%%%%%%%%%%%%%%%%%%%%%%%%%%%%%%%%%%%%%%%%%%%%%%%%%%%%%%%%%%%%%%%%%%%
\section{Data Availability}
The experimental data for this work have been made available online in Ref.~\cite{Experimental_Data_2026}

% SECTION VIII %%%%%%%%%%%%%%%%%%%%%%%%%%%%%%%%%%%%%%%%%%%%%%%%%%%%%%%%%%%%%%%%%%%%%%%%%%%%%%%%%%
\section{Acknowledgments}
We thank Ramona Wolf for helpful discussions and Stefano Pironio for suggesting multiple key generation bases analysis. M.~Malik, S.~Goel, B.~Ghosh, V.~Srivastav, and W.~McCutcheon acknowledge financial support from the European Research Council (ERC) Starting Grant PIQUaNT (950402), the UK Engineering and Physical Sciences Research Council (EPSRC) (EP/Z533208/1, EP/W003252/1), and the Royal Academy of Engineering Chair in Emerging Technologies programme (CiET-2223-112). G.~Murta and M.~Mothsara acknowledge funding from the Austrian Research Promotion Agency (FFG) through the Project NSPT-QKD FO999915265. Colormaps for some figures in this article are adopted from~\cite{crameri_fabio_2021_5501399}. 

\newpage 
\bibliography{bibliography}

\newpage
\onecolumngrid
\appendix
\newpage

% APPENDIX A %%%%%%%%%%%%%%%%%%%%%%%%%%%%%%%%%%%%%%%%%%%%%%%%%%%%%%%%%%%%%%%%%%%%%%%%%%%%%%%%%%
\section{Observed steering functional $\beta_{\mathrm{obs}}$ for the extra-outcome strategy} \label{analytical_bound_derivation}
To model Alice's detector inefficiencies, the measurement operators on her side are given by
\begin{equation}
    M_{a|x}^{(\eta,\rm{eo})} = 
    \begin{cases}
        \eta \, M_{a|x} & \text{for } a = 0, \dots, d-1 \\
        (1 - \eta) \, \mathds{I} & \text{for } a = \varnothing\,, \label{Max_eo}
    \end{cases}
\end{equation}
if one treats no-click events as an extra outcome, or
\begin{equation}
    M_{a|x}^{(\eta,\rm{rand})} = 
    \begin{cases}
        \eta \, M_{a|x} +
        (1 - \eta) \, \mathds{I}/d & \text{for }a = 0, \dots, d-1\,, \label{Max_rand}
    \end{cases}
\end{equation}
in case of a random assignment strategy, or 
\begin{equation}
    M_{a|x}^{(\eta,\rm{det})} = 
    \begin{cases}
        \eta \, M_{a|x} +(1 - \eta) \, \mathds{I} & \text{for } a = 0 \\
        \eta \, M_{a|x}  & \text{for } a = 1, \dots, d-1, \label{Max_det}
    \end{cases}
\end{equation}
for a deterministic assignment strategy. Here $M_{a|x}$ are the projective MUB measurement operators, $\varnothing$ denotes the extra no-click outcome, and $\eta$ denotes the detection efficiency of Alice's detectors.

\begin{figure}[H]
\centering
\includegraphics[scale=0.475, keepaspectratio]{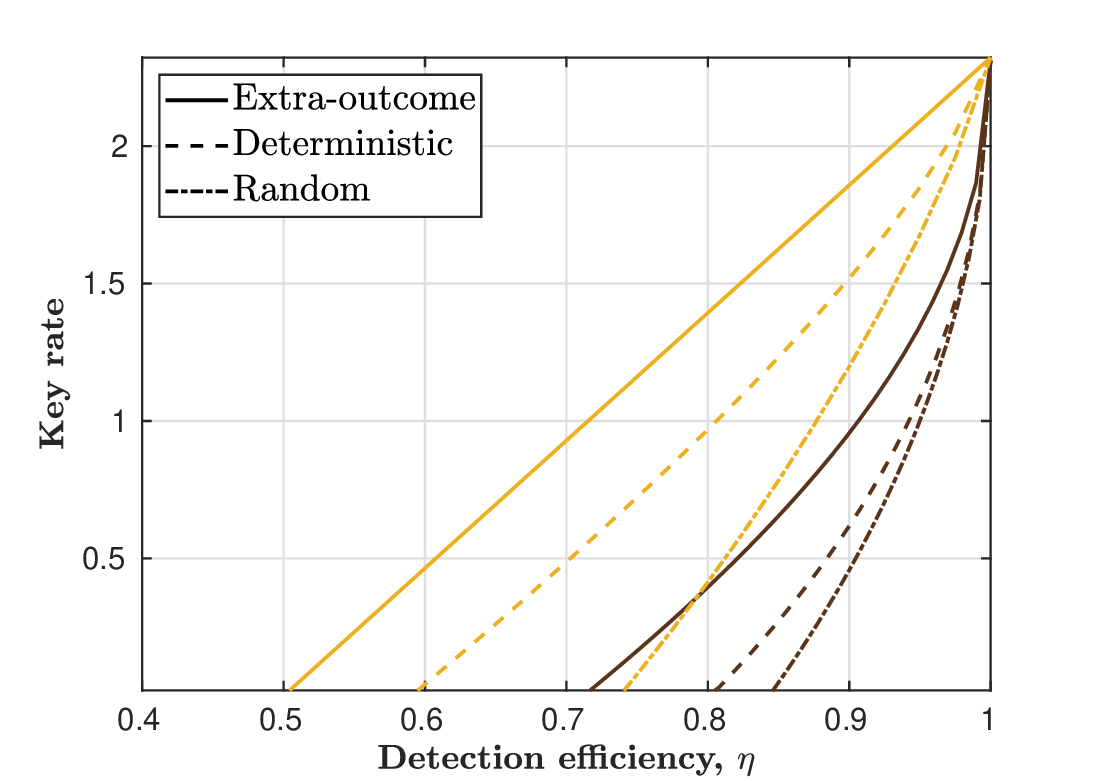}
\caption{{\bf Key rate as a function of detection efficiency for different loss treatments (dimension $d=5$).} {\bf\color{yellow}yellow} curves correspond to Protocol~\ref{protIa} (two-basis), while {\bf\color{brown}brown} curves denote \ref{protIb} (spot-checking). Solid lines correspond to the extra-outcome strategy for no-click events, while dashed and dash-dot lines denote deterministic and random assignments of no-click events, respectively. Visibility is assumed to be ideal ($\nu=1$).}
\label{loss_model_comparison}
\end{figure}

As an illustration, Fig.~\ref{loss_model_comparison} shows the key rate as a function of the detection efficiency $\eta$ for Protocol~\ref{protIa} and~\ref{protIb} in dimension $d=5$, comparing the three loss-treatment strategies. While the figure explicitly displays only the case $d=5$, we observe the same qualitative behavior for all dimensions considered, with the extra-outcome strategy consistently providing the best performance. Consequently, we adopt this strategy throughout the paper.\\

Below, we derive the expression for the observed steering functional $\beta_{\rm{obs}}$ for the extra-outcome strategy, to be used as an input to the SDPs in Eqs.~(\ref{eq:sdp_spotchecking}) and (\ref{eq:sdp_multikeybases}) for Protocols~\ref{protIb} and~\ref{protIc}. The corresponding assemblage obtained by applying Alice's measurement operators (Eq.~(\ref{Max_eo})) to the d-dimensional isotropic state subjected to depolarizing noise is given by
\begin{equation}
    \sigma_{a|x} = 
    \begin{cases}
        \frac{\eta \nu}{d} M_{a|x}^{\top}+\eta(1-\nu) \frac{\mathds{I}}{d^2} & \text{for } a=0,\dots,d-1 \\
        \left(1-\eta\right) \frac{\mathds{I}}{d} & \text{for }  a=\varnothing\,. \label{sigma_ax_B}
    \end{cases}
\end{equation}
Using this assemblage, we now evaluate the steering functional $\beta$ introduced in Eq.~(\ref{S.I.}):
\begin{equation}
    \beta = \mathrm{tr} \sum_{a,x} N_{b=a|y=x} \sigma_{a|x} \leq \beta^{\mathrm{LHS}}\,, \label{beta_obs}
\end{equation} 
Following the loss-tolerant steering inequality construction of~\cite{loss-tolerant_SI_Paul}, we use the following operators (applied to Bob's system) for the inequality
\begin{equation}
    N_{b|y}= 
    \begin{cases}
        N_{b \mid y} & \text { for } b=0, \dots, d-1 \\ 
        \alpha \mathds{I} & \text { for } b=\varnothing\,,
    \end{cases} \label{Nby}
\end{equation}
where $\{N_{b \mid y}\}_b$ denote Bob's measurements, which we consider to be MUBs, and $\alpha$ denotes the maximal overlap between any two measurements, which for MUBs equals $1/\sqrt{d}$. Substituting Eqs.~(\ref{sigma_ax_B}) and (\ref{Nby}) in Eq.~(\ref{beta_obs}), we obtain
\begin{equation}
    \begin{aligned}
            \beta_{\mathrm{obs}} &=  \mathrm{tr} \sum_{a, x} N_{b=a|y=x} \sigma_{a|x} +  \mathrm{tr} \sum_x N_{\varnothing \mid x} \sigma_{\varnothing \mid x} \\
            &=\eta \:\mathrm{tr} \sum_{a, x} N_{b=a|y=x}^{\top} \left(\frac{\nu}{d} M_{a|x}^{\top}+(1-\nu) \frac{\mathds{I}}{d^2}\right)+(1-\eta) \:\mathrm{tr} \sum_x (\alpha\mathds{I}) \frac{\mathds{I}}{d} \\
            &= m\eta\left(\nu+\frac{1-\nu}{d}\right) + \frac{m(1-\eta)}{\sqrt{d}}\,. \\ 
        \label{analytical_bound}
    \end{aligned}
\end{equation}

% APPENDIX B %%%%%%%%%%%%%%%%%%%%%%%%%%%%%%%%%%%%%%%%%%%%%%%%%%%%%%%%%%%%%%%%%%%%%%%%%%%%%%%%%%
\section{Why postselection may result in security loopholes in 1sDI-QKD} 
\label{EUR_analytical}

In~\cite{branciard2012one}, Branciard \textit{et al.} propose an argument suggesting that one may use only the rounds in which Alice obtains an outcome to form the key, i.e., by post-selecting on Alice’s loss events.

In this section, we revisit the argument of~\cite{branciard2012one} and identify the subtle point at which it breaks down. The reasoning in~\cite{branciard2012one} proceeds as follows: the smooth min-entropy of Bob's full string $B$ is related to the smooth min-entropy of Bob’s post-selected string $B^{\mathrm{ps}}$ via
\begin{subequations}\label{hmin_branciard}
\begin{align} 
    H_{\min }^\epsilon(B|E,X=Y=0) &= H_{\min }^\epsilon (B^{\mathrm{ps}}, B^{\mathrm{dis}}|E,X=Y=0)\label{psInfo}\\
    &\leq H_{\min }^\epsilon (B^{\mathrm{ps}}| B^{\mathrm{dis}}E,X=Y=0) + \log_2|B_1^{\mathrm{dis}}|\label{chainrule}\\
    &\leq H_{\min }^\epsilon (B^{\mathrm{ps}}|E,X=Y=0) + \log_2|B_1^{\mathrm{dis}}|\label{final_data_pro}
\end{align}
\end{subequations}Here, $B^{\mathrm{ps}}$ and $B^{\mathrm{dis}}$ denote the postselected (detection events) and discarded (no detection events) bit strings, where a chain rule for the smooth min-entropy is used in Eq.~(\ref{chainrule}), and the data-processing inequality is used in Eq.~(\ref{final_data_pro}). We emphasize that $B$ is already post-selected on Bob’s detection events, which is justified by the assumption of trusted devices on Bob’s side in the 1sDI setting.

The authors then proceed to apply the entropic uncertainty relation for smoothed entropies
\begin{equation}
H_{\min }^\epsilon(B|E,X=Y=0) \geq -\log_2 c - H_{\max }^\epsilon(B|A,X=Y=1)\,,\label{GEUR}
\end{equation}
where $c$ quantifies the maximum overlap between Bob's measurement bases. Substituting Eq.~(\ref{GEUR}) into Eq.~(\ref{final_data_pro}) yields
\begin{equation}
    H_{\min }^\epsilon(B^{\mathrm{ps}}|E,X=Y=0) \geq -\log_2 c - H_{\max }^\epsilon(B|A,X=Y=1) - \log_2|B_1^{\mathrm{dis}}|. \\
\end{equation}

The issue with this analysis already arises in Eq.~\eqref{psInfo}, where relevant information accessible to the eavesdropper Eve is not made explicit. In fact, if we are able to split the string $B$ into the strings $B^{\mathrm{ps}}$ and $ B^{\mathrm{dis}}$, the labels of which rounds belong to each of these strings is information available to Eve. To make this explicit, we introduce a classical register $T$ that records this information (e.g., $T$ can be a bit string where $T_i=1$ if round $i$ is kept and $T_i=0$ if round $i$ is discarded). Therefore, in a protocol where Alice announces the discarded rounds, the relevant entropies we want to estimate should include Eve's knowledge about $T$, and  Eqs.\eqref{hmin_branciard} should be rewritten as
\begin{subequations}
\begin{align} 
    H_{\min }^\epsilon(B|E,T,X=Y=0) &= H_{\min }^\epsilon (B^{\mathrm{ps}}, B^{\mathrm{dis}}|E,T,X=Y=0)\label{psInfo2}\\
    &\leq H_{\min }^\epsilon (B^{\mathrm{ps}}| B^{\mathrm{dis}}E,T,X=Y=0) + \log_2|B_1^{\mathrm{dis}}|\label{chainrule2}\\
    &\leq H_{\min }^\epsilon (B^{\mathrm{ps}}|E,T,X=Y=0) + \log_2|B_1^{\mathrm{dis}}|\label{final_data_pro2}
\end{align}
\end{subequations}

Thus, in order to proceed with the argument of~\cite{branciard2012one}, we would in fact need to bound $H_{\min }^\epsilon(B|E,T,X=Y=0)$. However, the entropic uncertainty relation cannot be directly applied to this quantity without additional assumptions. This is precisely where the argument fails.

The underlying problem is that the post-selection is performed by Alice, whose measurement device is untrusted. Consequently, the variable $T$ may depend on Alice’s input $X$ (for instance, a malicious device could decide whether to produce an outcome or declare loss depending on $X$ in order to bias the observed statistics). In contrast, the entropic uncertainty relation applies to a fixed tripartite state $\rho_{ABE}$, relating the entropy of the measurement outcome of $B$ in one basis conditioned on $E$ to the entropy of the measurement outcome of $B$ in another basis conditioned on $A$. The additional register $T$ cannot, in general, be absorbed into $E$ or $A$, since it is generated after (and potentially correlated with) Alice’s measurement choice $X$. As a result, the dependence of $T$ on $X$ prevents a direct application of the EUR relation~(\ref{GEUR}).

This explicit argument also clarifies when the argument of~\cite{branciard2012one} is valid: it holds when $T$ is independent of Alice's measurement choice.  This corresponds precisely to assuming that losses are independent of Alice’s measurement input, i.e., assuming fair-sampling, which is a strong assumption in scenarios where Alice’s device is untrusted.

% APPENDIX C %%%%%%%%%%%%%%%%%%%%%%%%%%%%%%%%%%%%%%%%%%%%%%%%%%%%%%%%%%%%%%%%%%%%%%%%%%%%%%%%
\section{Guessing probabilities for \mbox{$d+1$}-basis protocols}
\label{app:guessing_prob}
To model Eve's side information, we consider the purification $\rho_{\mathrm{ABE}}$ of the bipartite state $\rho_{\mathrm{AB}}$ with $\mathrm{tr}_\mathrm{E}(\rho_{\mathrm{ABE}}) = \rho_{\mathrm{AB}}$, where system $E$ is held by an adversary, Eve.
\subsection{Spot-checking protocol (Protocol \ref{protIb})}\label{fixed_basis}
We begin by considering a scenario where Eve performs a measurement on her side information $M_e$ with outcomes $0,...,d-1$ in an attempt to guess Bob's measurement outcome. The resulting assemblage, conditioned on Alice’s and Eve’s measurement outcomes, is defined as
\begin{equation}
    \sigma_{a|x}^{e} = \mathrm{tr}_{AE} \left[\left( M_{a|x} \otimes \mathds{I}_B \otimes M_e \right) \rho_{\mathrm{ABE}} \right]\,,
    \label{assemblage_sc}
\end{equation}
Averaging over Eve's outcomes recovers the assemblage held by Bob, $\sigma_{a|x}=\sum_{e} \sigma_{a|x}^{e}$. Eve's guessing probability for fixed measurement basis settings \mbox{$Y=y^*$} and \mbox{$X=x^*$} can then be written as 
\begin{equation} \label{Pguess_BE_sc}
    \begin{aligned}
        P_{\mathrm {guess}}(B|E,X=x^*,Y=y^*) 
        &= \max_{\substack{M_e, \\ \rho_{\mathrm{ABE}}}} \sum_{b=0}^{d-1} \mathrm{tr} \left[ 
        \left( \mathds{I}_A \otimes N_{b|y^*} \otimes M_{e=b} \right) \rho_{\mathrm{ABE}} \right]\\
        & =  \max_{\substack{M_e, M_{a|x} \\ \rho_{\mathrm{ABE}}}} {\small \sum_{a=0}^{d}\sum_{b=0}^{d-1} \mathrm{tr} \left[ 
        \left( M_{a|x^*} \otimes N_{b|y^*} \otimes M_{e=b} \right) \rho_{\mathrm{ABE}} \right]}\\
        &= \max_{\{\sigma_{a|x}^{e}\}}  {\small \sum_{a=0}^{d}\sum_{b=0}^{d-1} \mathrm{tr} \left[ N_{b|y^*} \, \sigma_{a|x^*}^{e=b} \right]}\,,
    \end{aligned}
\end{equation}
where the last equality follows from Eq.~(\ref{assemblage_sc}). Thus, the optimization problem can be expressed entirely in terms of the assemblage $\{\sigma_{a|x}^{e}\}$.
\subsection{Multiple key-generation basis protocol (Protocol \ref{protIc})}\label{all_bases}
In this protocol, Alice and Bob choose their inputs randomly among $d+1$ measurement bases with probability distribution $p(x,y)$. Here we choose a uniform distribution of all the inputs, i.e., $p(x,y)=1/m$ for all $x,y$ $\in$ $\{0,\ldots,\mbox{$m-1$}\}$. Since Bob's measurement setting is publicly announced, Eve is allowed to adapt her measurement strategy accordingly, and performs a measurement $M_{e|z}$ with input $z=\{y\}$ and outcome $e$. We therefore define the assemblage conditioned on Alice's measurement outcome and Eve's measurement outcome for a given input $z$ as
\begin{equation}
    \sigma_{a|x}^{e,z} = \mathrm{tr}_{AE} \left[ 
    \left( M_{a|x} \otimes \mathds{I}_B \otimes M_{e|z} \right) \rho_{\mathrm{ABE}} \right]\,,
    \label{assemblage_mkb}
\end{equation}
where averaging over Eve's outcomes recovers the assemblage held by Bob, $\sigma_{a|x}=\sum_{e} \sigma_{a|x}^{e,z}$ for every $z$. The guessing probability of Eve, averaged over all Bob's measurement settings according to the distribution $p(y)$, is then given by
\begin{equation} \label{Pguess_BE_mkb}
    \begin{aligned}
        P_{\mathrm {guess}}(B|E,X,Y)
        &= \max_{\substack{\{M_{e|z}\}_y, \\ \rho_{\mathrm{ABE}}}} \sum_{y\in Y} p(y) \sum_{b=0}^{d-1} \mathrm{tr} \left[\left( \mathds{I}_A \otimes N_{b|y} \otimes M_{e=b|z=y} \right) \rho_{\mathrm{ABE}} \right]\\
        &=  \max_{\substack{\{M_{e|z}\}_y, M_{a|x} \\ \rho_{\mathrm{ABE}}}} \sum_{y\in Y} p(y) \sum_{a=0}^{d}\sum_{b=0}^{d-1} \mathrm{tr} \left[ 
        \left( M_{a|x^*} \otimes N_{b|y} \otimes M_{e=b|z=y} \right) \rho_{\mathrm{ABE}} \right]\\
       &= \max_{\{\sigma_{a|x}^{e,z}\}} \sum_{y\in Y} p(y) \sum_{a=0}^{d}\sum_{b=0}^{d-1} \mathrm{tr} \left[ N_{b|y} \, \sigma_{a|x^*}^{e=b,z=y} \right]\,,
    \end{aligned}
\end{equation}
where the last equality follows from Eq.~(\ref{assemblage_mkb}).

% APPENDIX D %%%%%%%%%%%%%%%%%%%%%%%%%%%%%%%%%%%%%%%%%%%%%%%%%%%%%%%%%%%%%%%%%%%%%%%%%%%%%%%%%%
\section{Steering detection w.r.t. reverse and direct reconciliation}
\label{steering_detection_wrt_direct_reverse}
Steerability is characterized by the absence of a local hidden state (LHS) model. To determine whether a given assemblage $\{\sigma_{a|x}\}_{a,x}$ is steerable, we employ the following semidefinite program (SDP) optimization from Ref.~\cite{cavalcanti2016quantum},
\begin{equation}
    \begin{aligned}
    \text { given } \quad & \left\{\sigma_{a \mid x}\right\}_{a, x},\{D(a \mid x, \lambda)\}_{\lambda} \\
    \max _{\left\{\sigma_\lambda\right\}} \quad & \mu \\
    \text { s.t. } \quad & \sum_{\lambda} D(a \mid x, \lambda) \sigma_{\lambda}=\sigma_{a \mid x} \quad \forall a, x \\
    & \sigma_\lambda \geqslant \mu \mathds{I} \quad \forall \lambda
    \label{detection_sdp}
    \end{aligned}
\end{equation}
Here, $D(a|x,\lambda)$ denotes the deterministic response function, where $a=\lambda(x)$, and $\lambda(\cdot)$ is a function from $\{0,\ldots,m-1\}$ to $\{0,\ldots,d-1\}$. $\sigma_{\lambda}$ are the optimization variables of the SDP, corresponding to the members of the LHS model satisfying $\sum_{\lambda}\operatorname{tr}(\sigma_{\lambda})=1$. A negative optimal value of $\mu$ certifies that no LHS model exists and therefore implies that the assemblage is steerable, whereas $\mu\ge0$ implies that the assemblage is compatible with an LHS model.

In the following, we restrict our analysis to the qubit case $d=2$. For the steering inequality in Eq.~(\ref{S.I.}) together with the considered MUBs, the corresponding LHS bound is
$\beta_{\mathrm{LHS}}\approx1.71$ by maximizing the dual of SDP (\ref{detection_sdp}) over all unsteerable assemblages. In the direct-reconciliation scenario, shown in Fig.~\ref{fig:PgAE_betaobs}, the threshold at which $P_{\mathrm{guess}}(A|E,X,Y)<1$ coincides with the LHS bound $\beta_{\mathrm{obs}}\approx 1.71$. But, interestingly, in the reverse reconciliation scenario, shown in Fig.~\ref{fig:PgBE_betaobs}, $P_{\mathrm{guess}}(B|E,X,Y)<1$ already appears at $\beta_{\mathrm{obs}} \approx 1.51$, well below the steering violation threshold, indicating that randomness is witnessed even before the considered steering inequality is violated. As discussed in the main text, this counterintuitive behavior can be understood by the fact that the secret key rate, and not the randomness, is a witness of entanglement. Consistently, we obtain a positive key rate only after the steering inequality is violated.

To further clarify this behavior, we apply the SDP in Eq.~(\ref{detection_sdp}) to the optimal assemblages obtained from the SDP for $P_{\mathrm{guess}}$ optimization. Interestingly, we find that the resulting optimal values of $\mu$, shown in Fig.~\ref{fig:mu_betaobs_PgBE}, become negative already at $\beta_{\mathrm{LHS}}\approx1.51$, demonstrating that the corresponding assemblages are indeed steerable despite not violating our considered steering inequality with MUBs.  

\begin{minipage}[t]{0.47\linewidth}
        \centering
        \includegraphics[width=\linewidth]{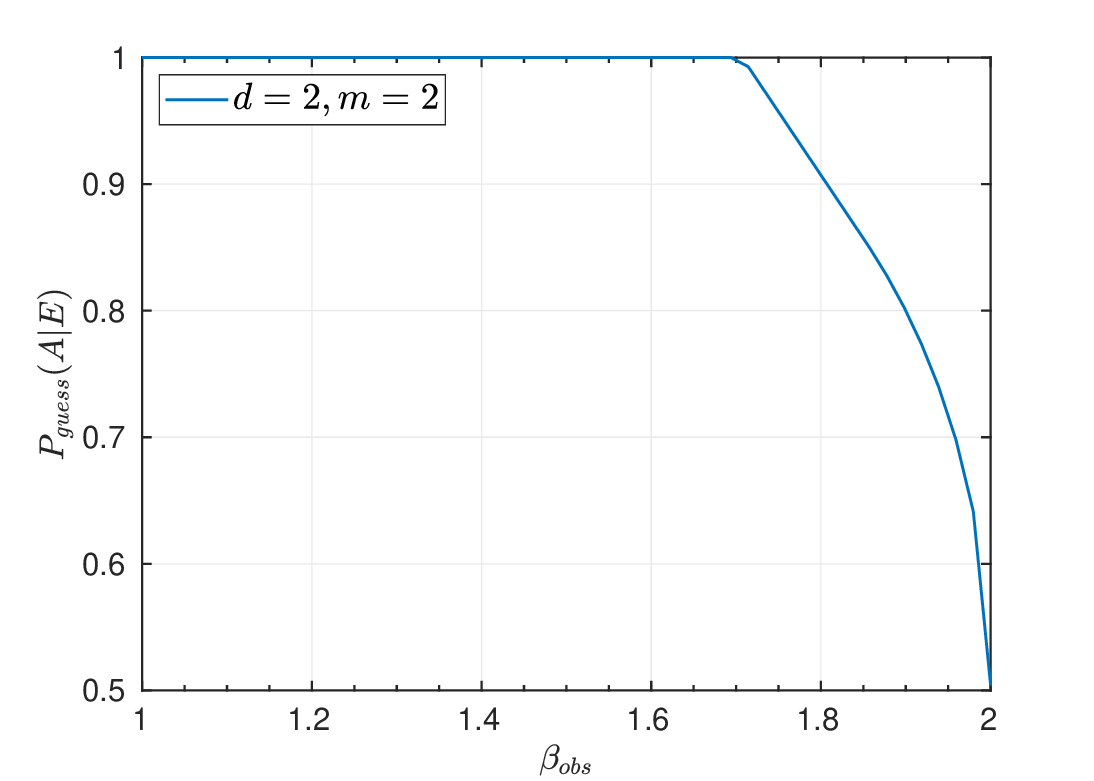}
        \captionof{figure}{Direct-reconciliation: Guessing probability $P_{\mathrm{guess}}(A|E)$ as a function of observed steering functional $\beta_{\mathrm{obs}}$}
        \label{fig:PgAE_betaobs}
\end{minipage}
    \hfill
\begin{minipage}[t]{0.47\linewidth}
        \centering
        \includegraphics[width=\linewidth]{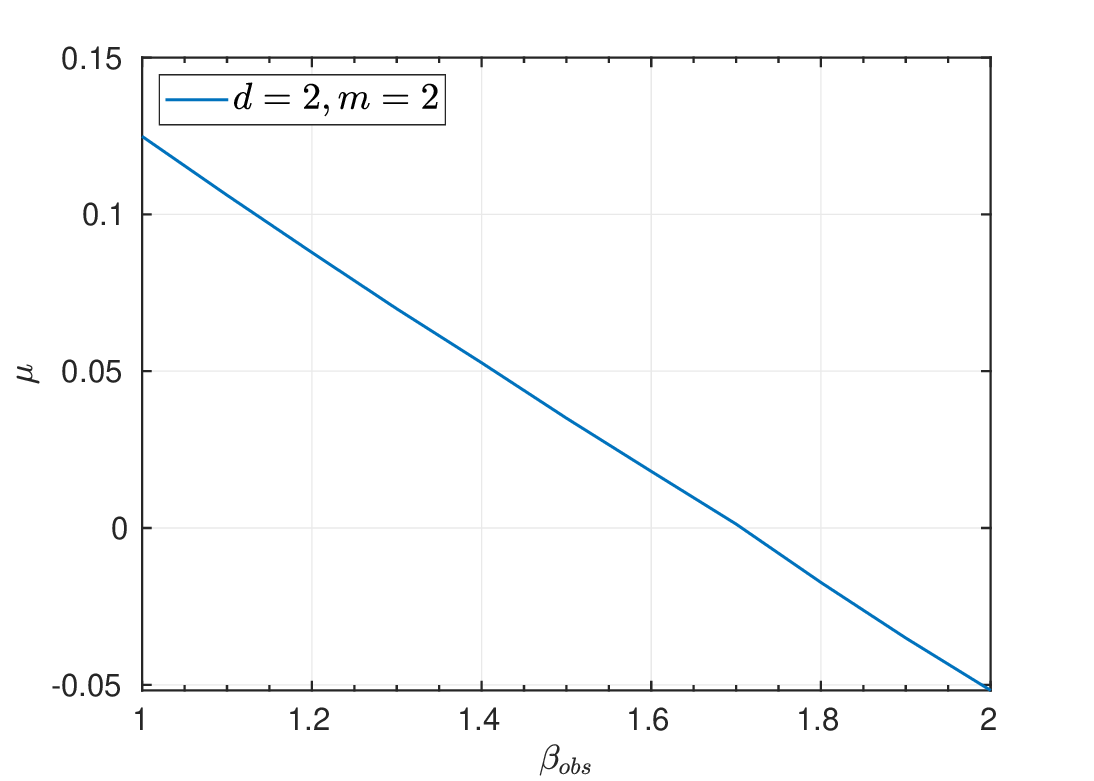}
        \captionof{figure}{Direct-reconciliation: Steering detection parameter $\mu$ as a function of observed steering functional $\beta_{\mathrm{obs}}$}
        \label{fig:mu_betaobs_PgAE}
\end{minipage}
\begin{minipage}[t]{0.47\linewidth}
        \centering
        \includegraphics[width=\linewidth]{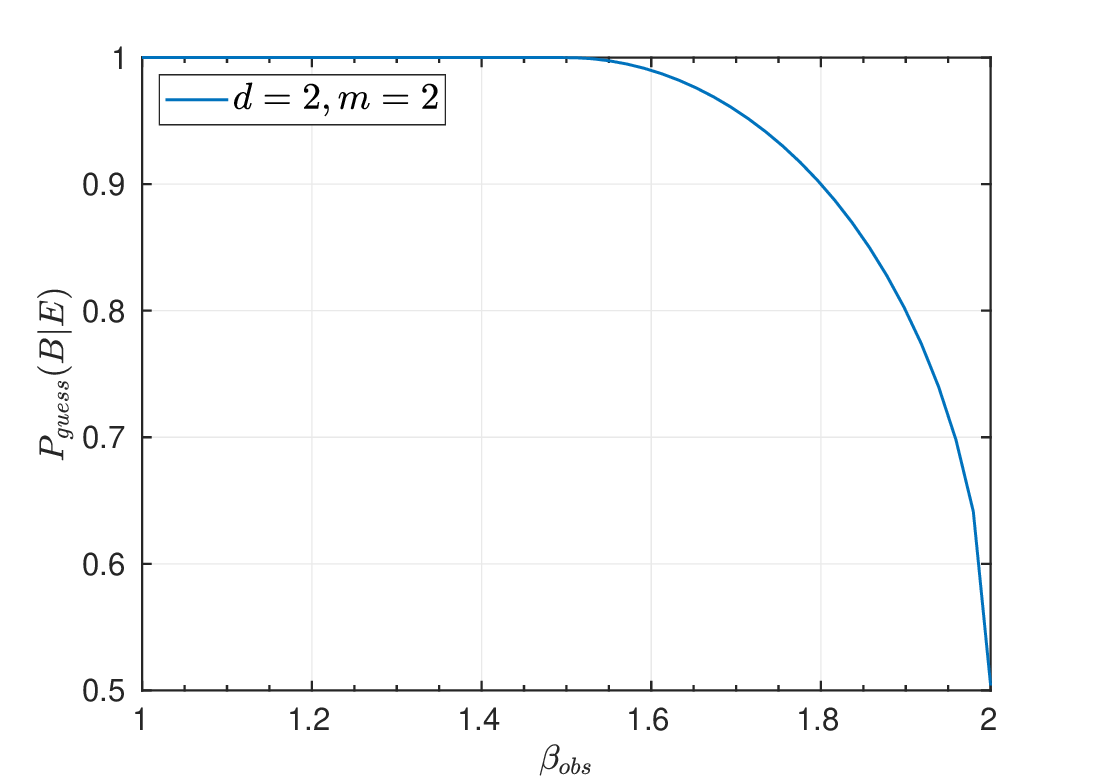}
        \captionof{figure}{Reverse-reconciliation: Guessing probability $P_{\mathrm{guess}}(B|E)$ as a function of observed steering functional $\beta_{\mathrm{obs}}$. }
        \label{fig:PgBE_betaobs}
\end{minipage}
    \hfill
\begin{minipage}[t]{0.47\linewidth}
        \centering
        \includegraphics[width=\linewidth]{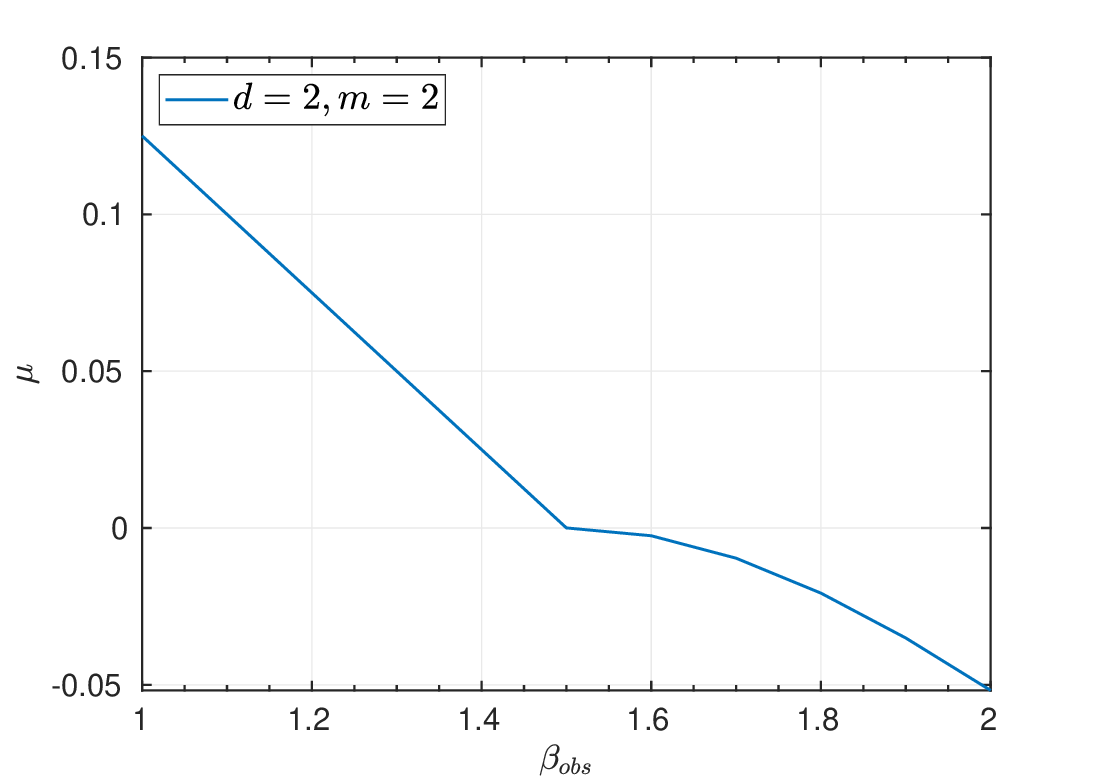}
        \captionof{figure}{Reverse-reconciliation: Steering detection parameter $\mu$ as a function of observed steering functional $\beta_{\mathrm{obs}}$}
        \label{fig:mu_betaobs_PgBE}
\end{minipage}

% APPENDIX E %%%%%%%%%%%%%%%%%%%%%%%%%%%%%%%%%%%%%%%%%%%%%%%%%%%%%%%%%%%%%%%%%%%%%%%%%%%%%%%%%%
\section{Information reconciliation} \label{error_correction}
As shown in Fig.~\ref{loss_model_comparison}, the extra-outcome strategy performs the best to account for no-click events. We assign an additional outcome $\varnothing$ to represent a no-click event, denoted by $\perp$.  
The conditional entropy of Bob's outcomes conditioned on Alice's outcomes is given by
    \begin{align}
        H(B|A) &= - \sum_{a\in\{0,\ldots,d-1,\perp\}}\sum_{b\in\{0,\ldots,d-1\}} p(b,a) \log p(b|a) \\
        &= - \sum_{a\in\{0,\ldots,d-1,\perp\}}\sum_{b\in\{0,\ldots,d-1\}} p(b,a) \log \frac{p(b,a)}{p(a)}\,.
        \label{H_AB}
    \end{align}
\noindent The distributed state is a noisy $d$-dimensional maximally entangled state with visibility $\nu$, and Alice’s detection efficiency $\eta$, given by
\begin{equation}
\rho_{\mathrm{AB}} = \eta \left[ \nu|\Phi^+_d\rangle\langle\Phi^+_d| + (1-\nu)\tfrac{\mathds{I}}{d^2} \right] 
+ (1-\eta) \left[|\!\perp\rangle\langle\perp\!|\otimes\tfrac{\mathds{I}}{d}\right],
\end{equation}
The resulting joint probability distribution $p(a,b)$ is as follows
\begin{itemize}
    \item $p(a=b; a=b\neq\perp) = \tfrac{\eta}{d}\!\left(\nu+\tfrac{1-\nu}{d}\right)$
    \item $p(a\neq b; a=\perp ) = \tfrac{1-\eta}{d}$
    \item $p(a\neq b; a\neq b\neq\perp) = \tfrac{\eta}{d}\left(\tfrac{1-\nu}{d}\right)$
\end{itemize}
And the marginal distribution $p(a)$ over Alice’s outcome is therefore
\begin{equation}
p(a)= \begin{cases}
        \eta / d, & \text{for } a = 0, \dots, d-1\\
        1-\eta, & \text{for } a = \perp
      \end{cases}
\end{equation}
Substituting these probabilities into Eq.~(\ref{H_AB}), we obtain 
\begin{equation}
    H(B|A) = -\eta\left(\nu + \frac{1-\nu}{d}\right)
    \log_2\!\left(\nu + \frac{1-\nu}{d}\right)
    + (1-\eta)\log_2 (d) - \eta(d-1)\left(\frac{1-\nu}{d}\right)\log_2\!\left(\frac{1-\nu}{d}\right)
    \label{eq:model_inf_rec}
\end{equation}
which quantifies the associated leakage during reverse reconciliation.

%Appendix F%%%%%%%%%%%%%%%%%%%%%%%%%%%%%%%%%%%%%%%%%%%%%%%%
\section{Experimental details}
\label{appendix_experimental_details}
\subsection{Details on the experimental setup}
\label{exp_app_details}
Telecom-band entangled photon pairs are generated via type-II spontaneous parametric down conversion (SPDC), pumped by a $775~\mathrm{nm}$ continuous-wave (CW) laser with an average power of $400~\mathrm{mW}$. The Gaussian pump beam is shaped using a telescope to optimize the generation of high-dimensional spatial entanglement. The residual pump is removed after down conversion using a dichroic mirror, followed by a long-pass filter.

The filtered biphoton field is imaged through a lens and spatially separated using a polarizing beam displacer. The two vertically displaced photons are then incident on a multi-plane light converter (MPLC), whose first reflective plane is placed at the back focal plane of the preceding lens. The MPLC is realized using a spatial light modulator (SLM, Holoeye PLUTO-2.1-TELECO-142) and a parallel mirror, aligned such that both photons reflect from the SLM three times, with a free-space propagation distance of $65~\mathrm{mm}$ between successive planes.

Following the MPLC transformation, the photons are coupled into two single-mode cores of a multi-core fiber placed at the back focal plane of the coupling lens, after transversal recombination on another polarizing beam-displacer. Each core is connected to a superconducting nanowire single-photon detector (SNSPD), and the resulting detection events are correlated using a Swabian Time Tagger Ultra with a coincidence window of $0.6~\mathrm{ns}$. 

For this proof-of-principle demonstration, we measure the joint-detection statistics by moving a single detector for each party and recording clicks for each pair of outcomes. 
The measured statistics correspond exactly to those obtained when multiple detectors are placed in parallel to record the outcomes under the fair-sampling assumption.
A natural next step is to use an array of efficient detectors~\cite{fleming2025high}, which would allow all outcomes to be detected in parallel and thereby remove this assumption.
%%%%%%%%%%%%%%%%%%%%%%%%%%%%%%%%%%%%%%%%%%%%%%%%%%%%%%%%%%%%%
\subsection{Data Processing}
\label{data_processing_details}
Given coincidence matrices $C^x_{a,b}$ and Bob's singles matrices $S^x_{a,b}$, for measurement setting $x$, Alice's outcome $a$, and Bob's outcome $b$, obtained from the experiment, we would like to construct an effective description that allows Bob to perform multi-outcome measurements, despite the losses on Alice's side.
Experimentally, Bob's singles are observed to be approximately invariant under the change of Alice's outcome, i.e., $S^x_{a,b} \approx S^x_{a',b}$. We then define Bob's exclusive singles as the events (averaged over Alice's outcomes) in which Bob records an outcome while Alice registers no clicks, ${\tilde S}^x_{b} = \tfrac{1}{d}\sum_a (S-C)^x_{a,b}$.

We now normalize the observed events. Let's define the total coincidence counts in basis ($x$),
$N_C^x = \sum_{a,b} C^x_{a,b}$, and the total exclusive singles counts, $N_S^x = \sum_b \tilde S^x_b$. The total number of detected events in basis $x$ is therefore
$N^x = N_C^x + N_S^x$. Thus, the normalized full-event, probability distribution over Alice’s outcomes ($a\in{1,\dots,d,\varnothing}$) and Bob’s outcomes ($b\in{1,\dots,d}$) is defined as follows. 
\begin{align}
    \tilde C^x_{a = \{1,...,d, \varnothing\},b = \{1,...,d\}} =   \begin{cases}
    C^x_{ab} /N^x & a=1,...,d\\
    \tilde S^x_b /N^x & a=\varnothing
  \end{cases}\,.
  \label{eq:norm_events}
\end{align}
By construction, these normalized events satisfy
$\sum_{a,b}\tilde C^x_{ab}=1$.

%%%%%%%%%%%%%%%%%%%%%%%%%%%%%%%%%%%%%%%%%%%%%%%%%%%%%%%%%%%%%
\subsubsection{Experimentally observed visibilities}
\label{exp_obs_vis}
From the coincidence matrices $C^x_{a,b}$, we compute normalized sums of diagonals, $V$, i.e., a measure of the cross-talk, defined as follows.
\begin{equation}
    V = \frac{1}{m}\sum_x\frac{\sum_aC^x_{a,a}}{\sum_{a,b}C^x_{a,b}}\,.
    \label{eq:expt_vis}
\end{equation}
From these visibilities, the corresponding depolarized state visibilities can be calculated using  $\nu_\mathrm{exp} = \frac{V-\frac{1}{d}}{1-\frac{1}{d}}$.
%%%%%%%%%%%%%%%%%%%%%%%%%%%%%%%%%%%%%%%%%%%%%%%%%%%%%%%%%%%%%
\subsubsection{Steering functional}
\label{expt_steering_data}
From the probabilities in Eq. (\ref{eq:norm_events}), the experimentally observed steering functional can be expressed as,

\begin{align}
    \beta_{\mathrm{obs}} = \sum_x \biggl( \sum_{a=1,...,d} \tilde C^x_{aa} + \alpha \sum_{b=1,...,d} \tilde C^x_{\varnothing,b} \biggr)\,.
    \label{eqs:betaObservedData}
\end{align}

The sum over $b$ in the second term is a consequence of Bob wanting to realize $N_{b=\varnothing|y=x}$ (Eq. (\ref{SIloss})), which is equivalent to summing over a complete POVM. Alternatively, $\beta_{\mathrm{obs}}$ can also be estimated by substituting experimentally obtained visibilities $\nu_{\mathrm{exp}}$ and estimated one-sided efficiencies $\eta_{\mathrm{exp}}$ (see Eq. (\ref{eqs:etaData}) below) into the model of a bipartite state subject to depolarizing noise and loss in Eq. (\ref{analytical_bound}). 
Tab.~\ref{tab:models_of_beta} below is a comparison between the values of $\beta_\mathrm{obs}$ for all investigated dimensions, showing that the values estimated using Eq.~(\ref{analytical_bound}) are in close agreement with the values computed from Eq.~\eqref{eqs:betaObservedData}.
\begin{table}[ht]
\centering
\caption{Observed steering violation estimated directly from the experimental data and from the model of depolarizing noise using the experimentally obtained visibilities $\nu_{\mathrm{exp}}$ and estimated one-sided efficiencies $\eta_{\mathrm{exp}}$.}
\begin{tabular}{ccc}
\hline
$d$ & $\beta_\mathrm{obs}$ from Eq. (\ref{eqs:betaObservedData}) & $\beta_\mathrm{obs}$ from Eq. (\ref{analytical_bound}) \\
\hline
2  & 2.143275 & 2.143290 \\
3  & 2.351195 & 2.351178 \\
5  & 2.736871 & 2.736695 \\
7  & 3.093733 & 3.093201 \\
9  & 3.385317 & 3.384598 \\
11 & 3.668987 & 3.668194 \\

\hline
\end{tabular}
\label{tab:models_of_beta}
\end{table}

%%%%%%%%%%%%%%%%%%%%%%%%%%%%%%%%%%%%%%%%%%%%%%%%%%%%%%%%%%%%%
\subsection{Handling experimental losses}
\label{handling_losses}
In the following sections, we account for losses in our system.

\subsubsection{Accounting for low experimentally measured one-sided detection efficiencies}
\label{simulating_eta}

\begin{figure}[H]
\centering
\begin{subfigure}{0.48\textwidth}
    \centering
    \includegraphics[width=\linewidth]{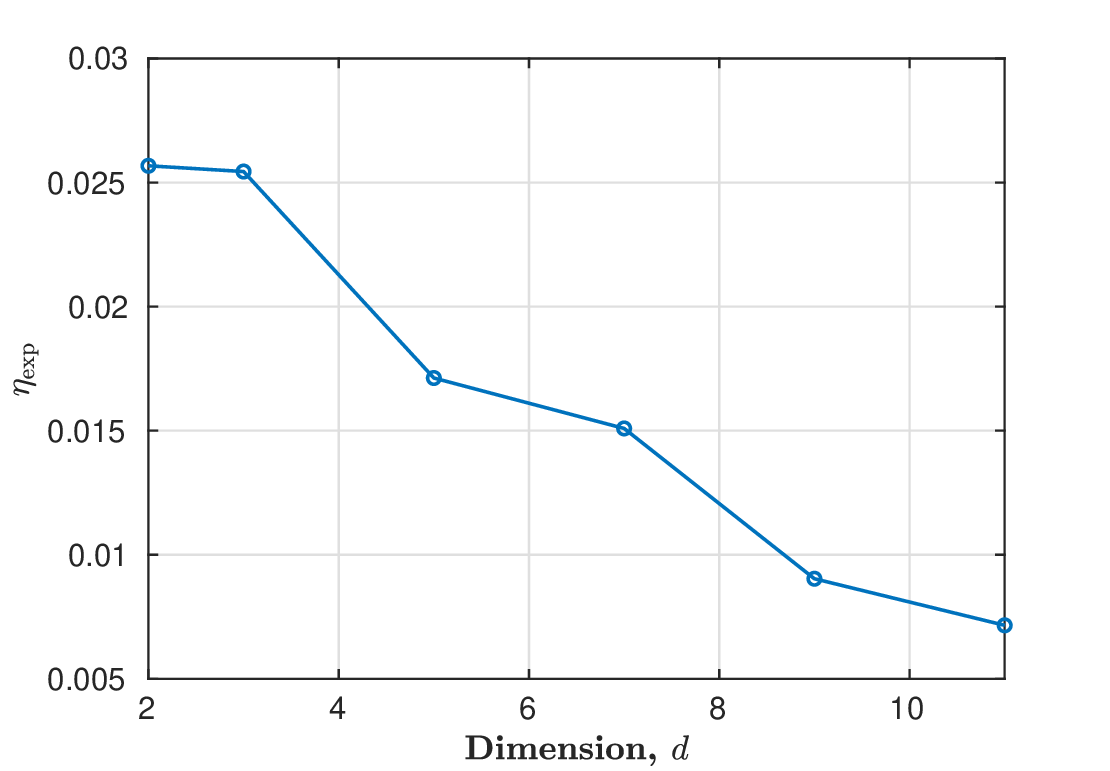}
    \caption{Experimentally measured one-sided detection efficiency $\eta_\mathrm{exp}$ (Eq. (\ref{eqs:etaData})) vs dimension $d$. Losses in the system, sources of which are discussed in Tab. (\ref{tab:appendix_efficiencies}), result in these sub-optimal efficiencies.}
    \label{exp_heralding_effi}
\end{subfigure}
\hfill
\begin{subfigure}{0.48\textwidth}
    \centering
    \includegraphics[width=\linewidth]{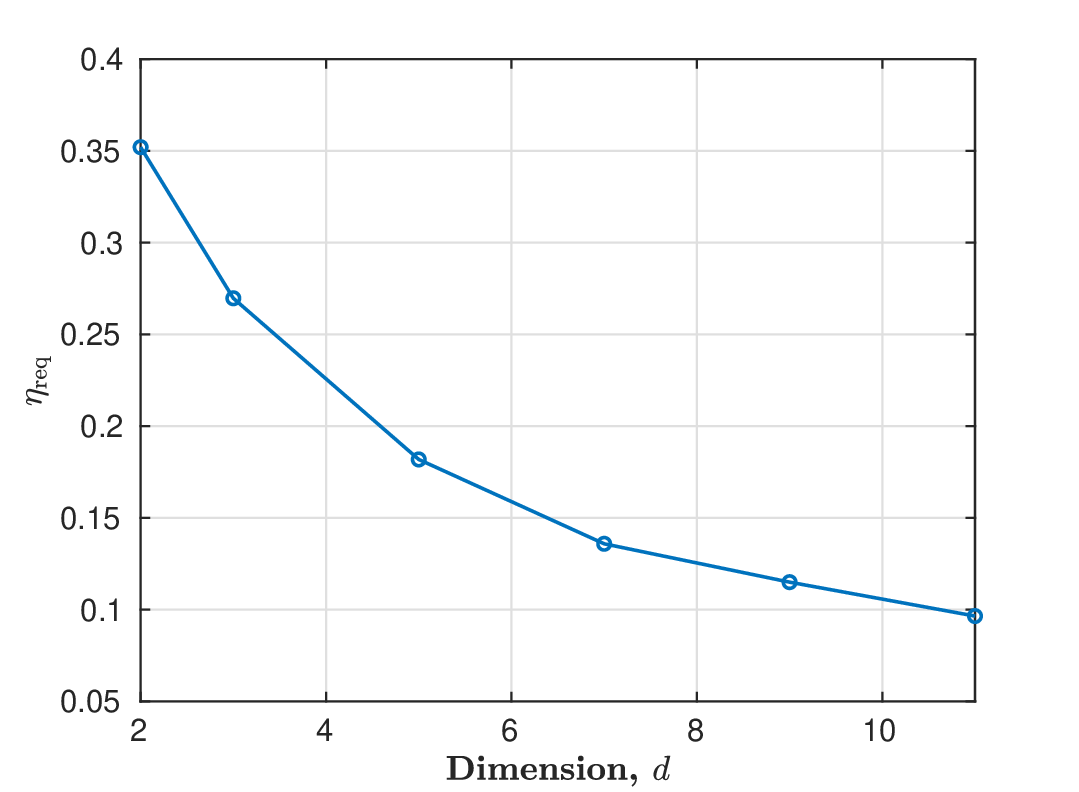}
    \caption{Required one-sided detection efficiency $\eta_\mathrm{req}$ (Eq. (\ref{eq:eta_req})) vs dimension $d$,
    required to allow the simulated steering functional to violate the LHS bound.
    }
    \label{sim_heralding_effi}
\end{subfigure}

\caption{Dimensional scaling of (a) one-sided detection efficiency obtained from the data and (b) one-sided detection efficiency required for the data to exhibit a steering violation, given the experimentally measured visibilities.}
\label{data_heralding_effi}

\end{figure}

In our data the average experimental one-sided detection efficiency $\eta_{\mathrm{exp}}$ is 
\begin{align}
    \eta_{\mathrm{exp}} = \frac{1}{m}\sum_x (N_C^x/N^x)\,.
    \label{eqs:etaData}
\end{align}
As discussed in the main text (also c.f. Fig.~\ref{exp_heralding_effi} below for a plot of $\eta_{\mathrm{exp}}$ vs $d$), owing to losses in the system, experimentally obtained values of $\eta_{\mathrm{exp}}$ are sub-optimal and currently below the bound $\eta>1/m$ required to even exhibit a detection loophole-free steering violation, a pre-requisite for loophole-free 1sDI-QKD.

In this section, we analyze the effect of suppressing different fractions of the single counts to simulate data that could be obtained with future higher-efficiency systems, whilst maintaining our experimentally measured visibilities. Suppressing singles by a factor $h$, resulting in singles rates $h {\tilde S}^x_b$, results in an improved simulated efficiency, 
\begin{align}
    \eta_{\mathrm{req}}(h) = \frac{1}{m}\sum_x \bigl(\frac{N_C^x}{N_C^x+hN_s^x}\bigr)\,
    \label{eq:eta_req}.
\end{align}
We then obtain a new probability distribution using Eq. (\ref{eq:norm_events}) from which we can extract the value of the steering functional $\beta_\mathrm{sim}$ (Eq. (\ref{eqs:betaObservedData}) applied to the simulated counts). We can then tune the efficiency from the raw experimentally observed efficiency $\eta_{\mathrm{exp}}$, to the fully post-selected case, $\eta\rightarrow 1$, via an intermediate regime where $\eta_{\mathrm{req}}\gtrsim 1/m$, to obtain the required one-sided detection efficiency that allows our data to exhibit a steering violation. We explore the critical efficiency we can tolerate to violate $\beta_{\mathrm{LHS}}$ with our experimental data, and plot this below.

As seen from Fig.~\ref{sim_heralding_effi}, for $d=7$, to violate $\beta_{\mathrm{LHS}}$ with experimentally measured visibilities, the required efficiency $\eta_{\mathrm{req}}\approx 0.14$. However, as seen in Fig  (\ref{exp_heralding_effi}), the experimental efficiency is $\eta_{\mathrm{exp}}\approx0.015$ for $d=7$. 
The estimated one-sided efficiency of the experimental setup implementing a $d=7$ operation is approximately 0.1067~(see Table~\ref{tab:appendix_efficiencies} below). We expect that the discrepancy between the measured experimental efficiency $\eta_{\mathrm{exp}}$ and the expected efficiency can be attributed to the experimental misalignments as well as the choice of spatial-mode bases in performing the measurements. Although macro-pixel bases lead to robust maximally entangled states, the effective local filtering of the underlying state's Schmidt modes leads to increased accidentals, thereby reducing one-sided efficiency~\cite{srivastav2022characterizing}. One can improve this efficiency in practice by transverse-spatially shaping the laser beam used to pump the SPDC process, effectively engineering a state that does not require local filtering~\cite{boucher2021engineering}.  

These considerations for achieving detection loophole-free steering serve as merely a necessary step towards fully 1sDI-QKD demonstrations, which, as this manuscript shows, require significantly more stringent efficiencies. In the next section, we list down losses associated with various components in the experiment, and discuss what attributes can be practically improved in the future. 

%%%%%%%%%%%%%%%%%%%%%%%%%%%%%%%%%%%%%%%%%%%%%%%%%%%%%%%%%%%%%
\subsubsection{Accounting for experimental losses}
\label{losses_explained}
The main instrumental efficiencies used to estimate the overall system efficiency are listed in Table~\ref{tab:appendix_efficiencies}. A primary source of loss in our experiment is the reflection and diffraction loss from each plane of the three-plane MPLC. In principle, this can be minimized by etching the phase masks on a reflective surface and gold-plating them, instead of using a commercially available SLM, to improve reflectivity~\cite{fang2021performance}. Additionally, each of the two beam displacers (Thorlabs BD40) used in our setup to split and recombine Alice's and Bob's paths (as detailed in section \ref{exp_app_details}) is lossy. They may be replaced with custom-built tunable beam displacers wherein the losses can be controlled~\cite{salazar2015tunable}. Furthermore, we lose photons via single-mode fiber coupling. This can be alleviated by utilizing an array of SNSPDs~\cite{fleming2025high}.
Lastly, quantum efficiencies of SNSPDs can be tuned to be higher than 0.9 by tuning their bias voltages. 

In addition to the aforementioned ways to improve instrumental efficiencies, better one-sided heralding efficiency can be achieved by sculpting asymmetric pixels out of the JTMA, as discussed in Appendix D of \cite{srivastav2022characterizing}. However, this may affect the visibility of our measurements.
\begin{table}[htbp]
    \centering
    \caption{Efficiencies of the main optical and detection components in the experimental setup.}
    \label{tab:appendix_efficiencies}
    \begin{tabular}{l c p{7.2cm}}
        \hline
        \textbf{Component} & \textbf{Efficiency} & \textbf{Description} \\
        \hline
        Three-plane MPLC 
        & $0.373$ 
        & Measured total diffraction and reflection efficiency of the three-plane MPLC. \\

        Efficiency of MPLC operation 
        & See Table.~\ref{tab:avg_mplc_sp}
        & MUB measurements implemented by MPLCs are typically imperfect.\\

        Beam displacer 
        & $0.85$ 
        & Efficiency of each beam displacer used to split and recombine Alice's and Bob's paths \\

        Single-mode fiber coupling 
        & $0.50$ 
        & Estimated coupling efficiency of the single-mode fibers used to couple photons to the detectors. \\

        SNSPD
        & $0.85$ 
        & Quantum efficiency of each superconducting nanowire single-photon detector at telecom wavelength [from Photon Spot Inc.]. \\
        \hline
    \end{tabular}
\end{table}

\begin{table}[ht]
\centering
\caption{Average expected efficiency over all MUB measurements for each party performed by the MPLC for each dimension $d$.}
\begin{tabular}{cc}
\hline
$d$ & Simulated Efficiency \\
\hline
2 & 0.966 \\
3 & 0.964 \\
5 & 0.939 \\
7 & 0.932 \\
9 & 0.919 \\
11 & 0.908 \\

\hline
\end{tabular}
\label{tab:avg_mplc_sp}
\end{table}
%%%%%%%%%%%%%%%%%%%%%%%%%%%%%%%%%%%%%%%%%%%%%%%%%%%%%%%%%%%%%
\subsubsection{Experimentally obtained information reconciliation}
\label{exp_inf_recon}
\begin{figure}[H]
\centering
\includegraphics[scale=0.475, keepaspectratio]{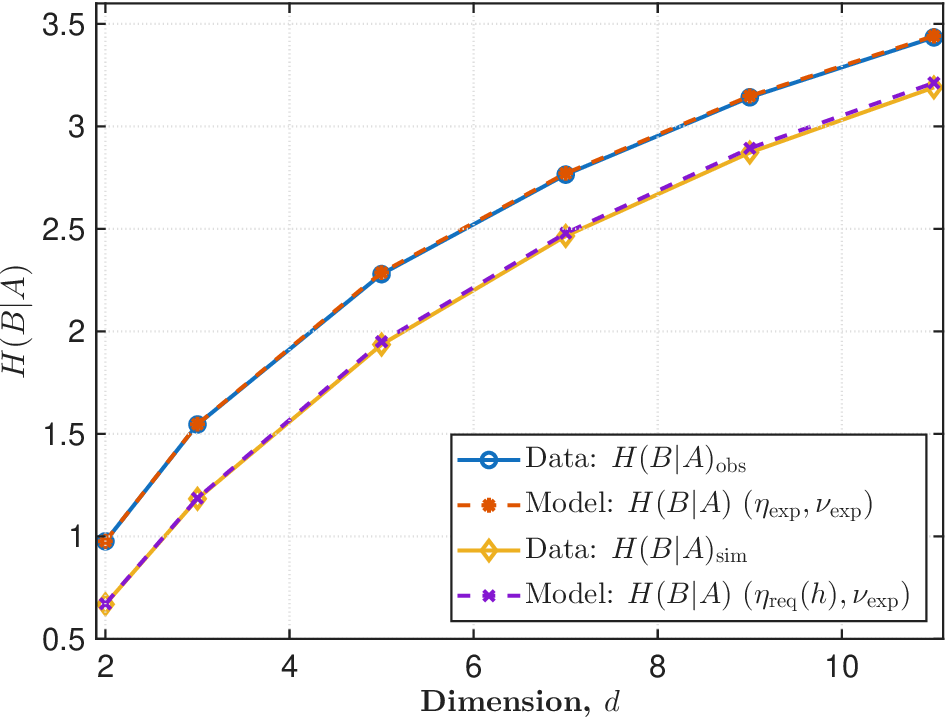}
\caption{Dimensional scaling of reverse information reconciliation. Solid lines show the information reconciliation obtained from the data using full and suppressed events (Eq. (\ref{eq:exp_inf_recon})). Overlapping dashed lines show the afore-mentioned quantities calculated by substituting experimentally measured  visibilities and efficiencies into the model of a bipartite state subject to depolarizing noise and loss (Eq. (\ref{eq:model_inf_rec})).} 
\label{inf_recon}
\end{figure}
Here we compute the experimentally estimated information reconciliation using the full date and also considering a suppression of Bob's exclusive singles, i.e., partially post-selected events.
Using the definitions of information reconciliation Eq. (\ref{H_AB}), and normalized events Eq. (\ref{eq:norm_events}), we have the experimentally observed information reconciliation,
\begin{equation}
    H(B|A)_\mathrm{obs} = -\frac{1}{m}\sum_{x,a,b}\tilde{C}^x_{a,b}\log_2\frac{\tilde{C}^x_{a,b}}{\sum_b\tilde{C}_{a,b}}
    \label{eq:exp_inf_recon}
\end{equation}
Similarly, a $H(B|A)_\mathrm{sim}$ can be used to represent simulated information reconciliation by suppressing some of Bob's exclusive events (c.f. subsection (\ref{simulating_eta})). Fig.~\ref{inf_recon} is a plot of experimentally observed $H(B|A)_\mathrm{obs}$ (full-event) and $H(B|A)_\mathrm{sim}$ (singles suppressed to obtain $\eta_{\mathrm{req}}=1/m+0.02$) vs $d$. As seen in the figure, experimentally obtained values closely overlap with the chosen model in Eq. (\ref{eq:model_inf_rec}). This suggests that experimentally obtained visibilities and efficiencies are uncorrelated.

%%%%%%%%%%%%%%%%%%%%%%%%%%%%%%%%%%%%%%%%%%%%%%%%%%%%%%%%%%%%%
\subsection{Correlations in All MUBs}
\label{corr_all_mub_data}
\begin{figure*}[b!]
    \centering
    \includegraphics[width=\textwidth]{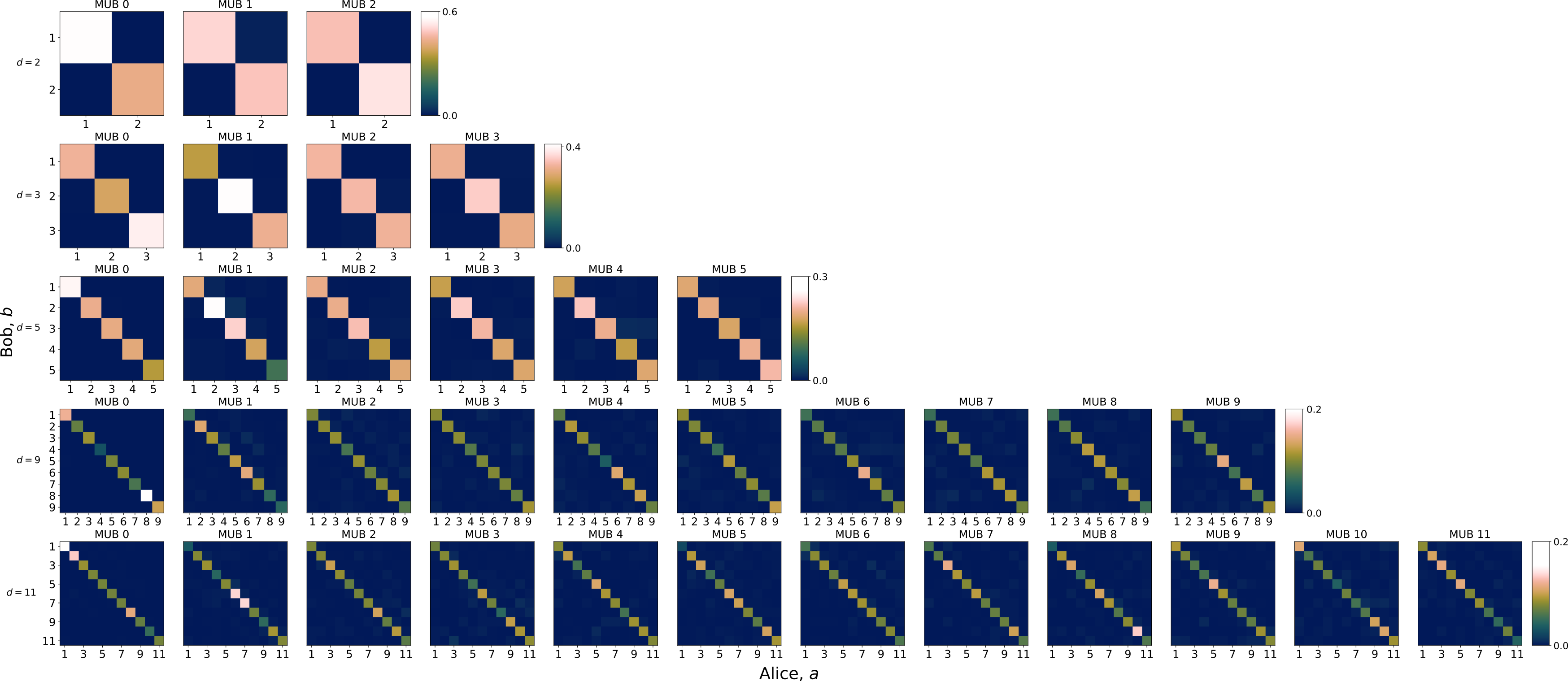}
    \caption{Normalized two-photon coincidence counts showing correlations across all MUBs in $d=2,3,5,9,11$. The data shows strong correlations and minimal cross-talk.}
    \label{all_mub_data}
\end{figure*}

Fig.~\ref{all_mub_data} shows measured two-photon correlations in all MUBs in $d=2,3,5,9,11$. For any dimension $d$, each of the $d+1$ plotted matrices represents the normalized two-photon coincidence matrices $\tilde{C}^x_{a,b}$ (see Eq.~\eqref{eq:norm_events}) obtained upon performing MUB measurements. A common color scale is used across all MUBs within a fixed dimension ($d$), with the color-bar ranging from zero to the largest value of $\tilde C^x_{ab}$ observed among all $d+1$ MUBs for that dimension. 
Below, in Tab.~\ref{tab:avg_vis_d}, we present the average measure of cross-talk $V$ (Eq.~\ref{eq:expt_vis}) for each investigated dimension. Depolarized state visibilities $\nu_{\mathrm{exp}}$ reported in Tab.~\ref{table1} of the main text are calculated using $V$. The normalized coincidence counts are independent of post-selection (any suppression of singles).

\begin{table}[H]
\centering
\caption{Average experimentally observed measure of cross-talk for each investigated dimension $d$.}
\begin{tabular}{cc}
\hline
$d$ & $V$ \\
\hline
2 & 0.992\\
3 & 0.988\\
5 & 0.967\\
7 & 0.953\\
9 & 0.901\\
11 & 0.884\\

\hline
\end{tabular}
\label{tab:avg_vis_d}
\end{table}

\end{document}